\documentclass[journal]{IEEEtran}
\usepackage{cite}

\ifCLASSINFOpdf
\else
\fi
\usepackage{amsmath}

\input{MySettings.tex}

\begin{document}
%
\title{Capacitively-coupled resonators for terahertz planar-Goubau-line filters}
%
%
%

\newcommand{\orcidauthorA}{0000-0002-9821-3199} 
\newcommand{\orcidauthorB}{0000-0002-5861-6655} 
\newcommand{\orcidauthorC}{0000-0001-9519-3527} 
\newcommand{\orcidauthorD}{0000-0002-8204-7894} 

\author{Juan~Cabello-S\'{a}nchez,~\IEEEmembership{Graduate Student Member,~IEEE,}
Vladimir~Drakinskiy,
Jan~Stake,~\IEEEmembership{Senior~Member,~IEEE,}
Helena~Rodilla,~\IEEEmembership{Senior~Member,~IEEE}
\thanks{Manuscript received 3\textsuperscript{rd} June, 2022; revised 17\textsuperscript{th} August, 2022; accepted 24\textsuperscript{th} October, 2022. This work was supported by the Swedish Research Council (Vetenskapsr\aa det) under grant 2020-05087.}
\thanks{Juan Cabello-S\'{a}nchez, Vladimir Drakinskiy, Jan Stake and Helena Rodilla are with the Terahertz and Millimetre Wave Laboratory, Chalmers University of Technology, SE-412 96 Gothenburg, Sweden. (e-mail: \mbox{juancab@chalmers.se}; \mbox{vladimir.drakinskiy@chalmers.se}; \mbox{jan.stake@chalmers.se}; \mbox{rodilla@chalmers.se})}
\thanks{Color versions of one or more of the figures in this article are available online at http://ieeexplore.ieee.org.}
\thanks{Digital Object Identifier \href{http://dx.doi.org/10.1109/TTHZ.2022.3220599}{10.1109/TTHZ.2022.3220599} }}%
%
%

\markboth{IEEE Transactions on Terahertz Science and Technology,~Vol.~XX No.~XX, XX~2022}%
{Cabello-S\'{a}nchez \MakeLowercase{\textit{et al.}}: Capacitively-coupled open-stub resonators for terahertz planar Goubau line filters}
%



\maketitle

\begin{abstract}
Low-loss planar Goubau lines show promising potential for terahertz applications.
However, a single-wire waveguide exhibits less design freedom than standard multi-conductor lines, which is a significant constraint for realizing standard components.
Existing filters for planar Goubau line lack clear design procedures preventing the synthesis of an arbitrary filter response.
In this work, we present a design for a bandpass/bandstop filter for planar Goubau line by periodically loading the line with capacitively coupled $\lambda/2$ resonators, which can be easily tuned by changing their electrical length.
The filter's working principle is explained by a proposed transmission-line model.
We designed and fabricated a passband filter centered at 0.9\,THz on a 10-$\mu$m silicon-membrane substrate and compared measurement results between 0.5\,THz and 1.1\,THz to electromagnetic simulations, showing excellent agreement in both $S_{11}$ and $S_{21}$.
The measured passband has an insertion loss of 7\,dB and a 3-dB bandwidth of 31\%.
Overall, the proposed filter design has good performance while having a simple design procedure.

\end{abstract}


%
\IEEEpeerreviewmaketitle


\section{Introduction}

The development of efficient radio-frequency technology and components at terahertz (THz) frequencies \cite{Mittleman2017} is essential for improving the performance and possibilities of THz applications, such as radio-astronomy \cite{Phillips1992}, security \cite{Appleby2007a}, medical applications \cite{Pickwell2006}, telecommunications \cite{Song2011}, pharmaceutical quality control \cite{Bawuah2021}, and biomolecular dynamics \cite{Acbas2014}.
When planar technology is required---for its good integration, ease of fabrication, and low cost---it is necessary to minimize the high power losses present at THz frequencies to have good circuit performance.
A fundamental way to minimize losses is to use power-efficient transmission lines.
Some of the most used metal planar waveguides at THz frequencies include coplanar waveguide \cite{Wen1969}, coplanar stripline \cite{Knorr1975}, microstrip \cite{Grieg1952}, and planar Goubau line (PGL) \textcolor{black}{\cite{Akalin2006}}.
As opposed to typical transmission lines based on multiple conductors \cite{Wu2021a}, the PGL is a single-conductor planar waveguide consisting of a metal strip on top of a dielectric material, and to the best of the authors' knowledge its properties were first mentioned in \textcolor{black}{\cite{Temnov2000,Hong2004}}.
The PGL is the planar version of the Goubau line \cite{Goubau1950, Goubau1951} whose propagation mode is similar to the Sommerfeld or Zenneck waves \cite{Barlow1953} with a field that decays exponentially in the transverse direction.
Similarly, the PGL can propagate a quasi-transverse-magnetic surface wave, which spreads radially from its single conductor, covering a large cross-section, and at the same time has higher field intensity at the edges of its conducting strip.
Its propagation conditions depend on the electrical thickness of its substrate \cite{Gacemi2013}, where electrically-thin substrates minimize losses and dispersion.
\textcolor{black}{The dispersion, effective refractive index, field confinement, and ohmic losses can be increased by adding periodic corrugations or grooves \cite{Rotmant1950}, a concept explored by Goubau in 1950 \cite{Goubau1950}, and sometimes referred to as spoof surface plasmon polariton waveguides \cite{Maier2006}.
Without corrugations,} the PGL has shown to have one of the highest power efficiencies for metal planar waveguides \cite{CabelloSanchez2018}, thus being a good candidate for circuit design at THz frequencies.

\begin{figure}[t!]
\centering%
\subfloat[]{%
\centering
\includegraphics[width=0.49\textwidth]{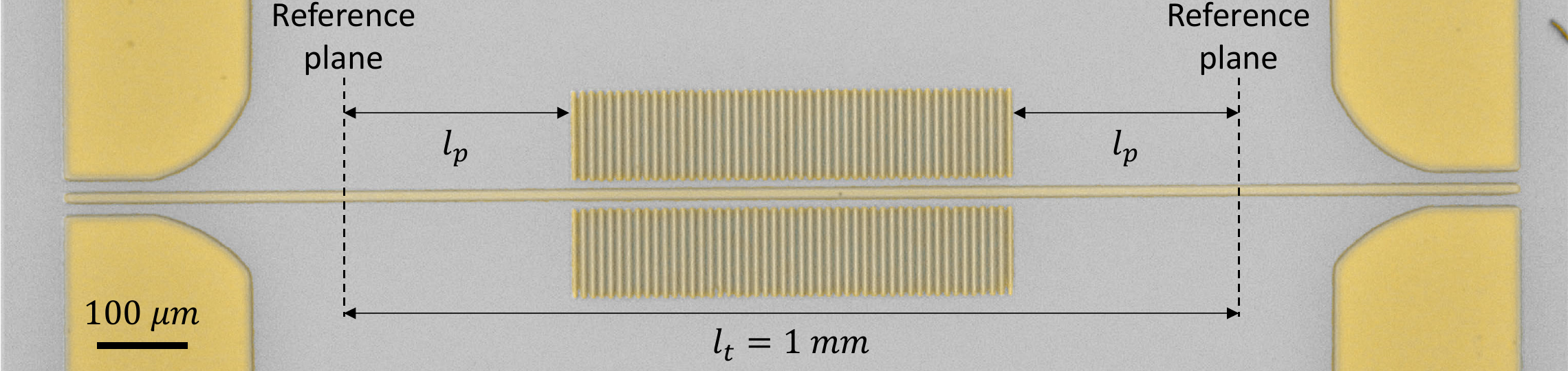}
\label{fig:filter_pic}
}%
\\[-0.1mm]%
\subfloat[]{%
\centering
\includegraphics[width=0.49\textwidth]{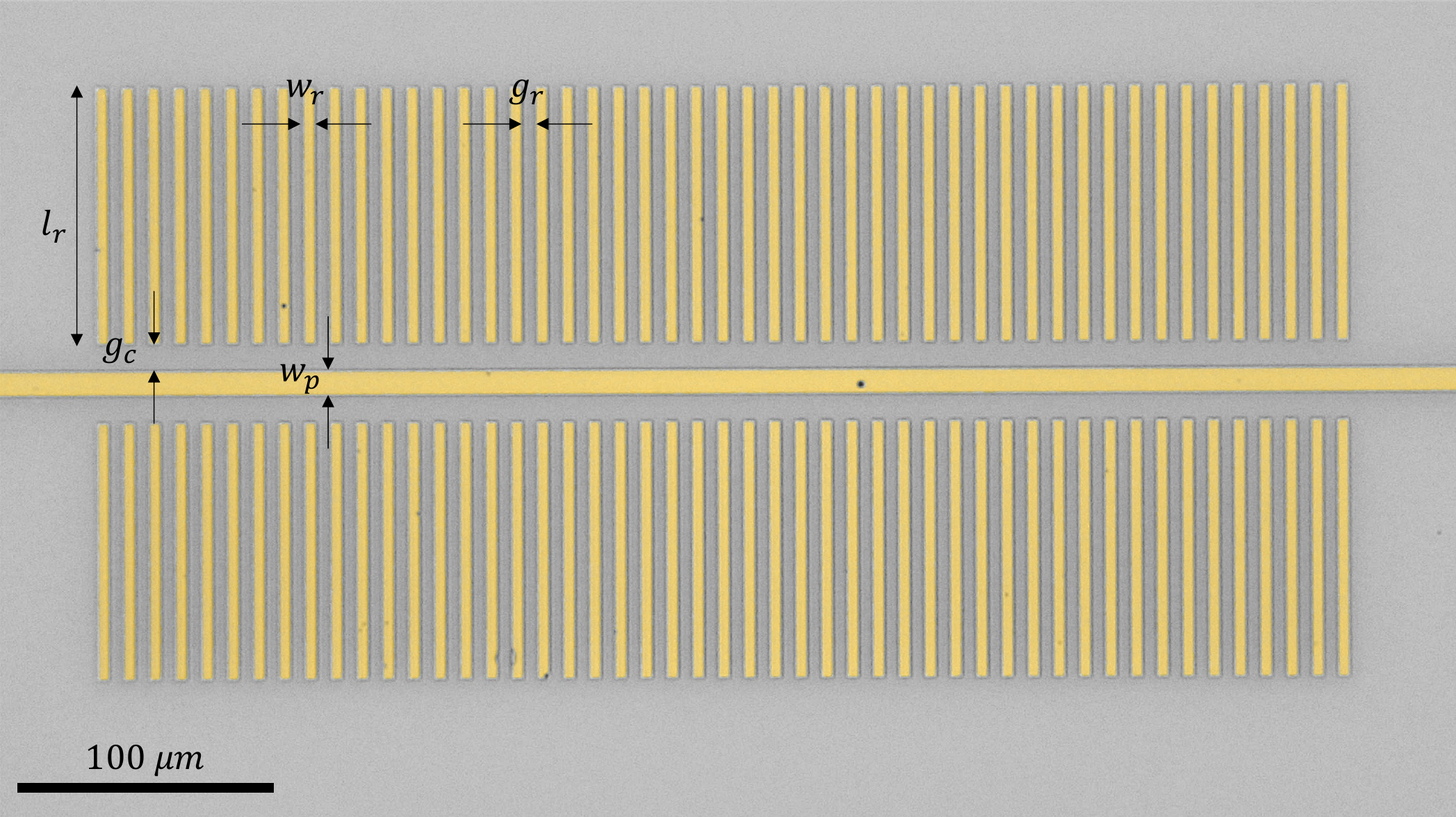}
\label{fig:filter_pic_zoom}
}%
\\[2.6mm]
\caption{False color micrographs of (a) the fabricated PGL filter with coplanar waveguide transitions on the sides, and (b) zoomed-in of the PGL filter with 49 resonator pairs, showing its dimension parameters.
The gold and the high-resistivity silicon membrane are shown in golden and grey colors, respectively.}
\label{fig:filter}
\end{figure}

Designing circuit elements for PGL remains challenging due to its lack of ground plane.
Despite this, several filtering circuit elements have been published for PGL, including stopband elements based on corrugations in the line \cite{Akalin2009} and resonators \cite{Dazhang2009,Treizebre2010,Chen2011,Horestani2013a}.
However, previously published resonating elements for PGL lack a clear design methodology that would allow them to synthesize any given filter response.
In \cite{Akalin2009}, it is unclear how to design line corrugations to produce an arbitrary stopband or if ring resonators can be used together to achieve a wider stopband.
The resonators used in \cite{Dazhang2009, Treizebre2010, Chen2011, Horestani2013a} can produce narrow stopbands, but they have not been tested by combining several resonators to achieve a broader bandwidth response.


This paper proposes a PGL filter based on periodic $\lambda/2$ resonators capacitively coupled to the PGL.
The resonant frequency of the resonators can be easily tuned by changing the length of the resonating lines.
The paper is organized as follows:
section \ref{sec:method} contains the method, divided in transmission-line model (section \ref{sec:model}), design of a bandpass filter at 0.9\,THz (section \ref{sec:design}), filter fabrication (section \ref{sec:fab}), and measurement setup 
(section \ref{sec:meas}).
Finally, in section \ref{sec:results}, we present the results, and the conclusion in section \ref{sec:concl}.



\section{Method} \label{sec:method}

\begin{figure}
\centering
\includegraphics[width=0.5\textwidth]{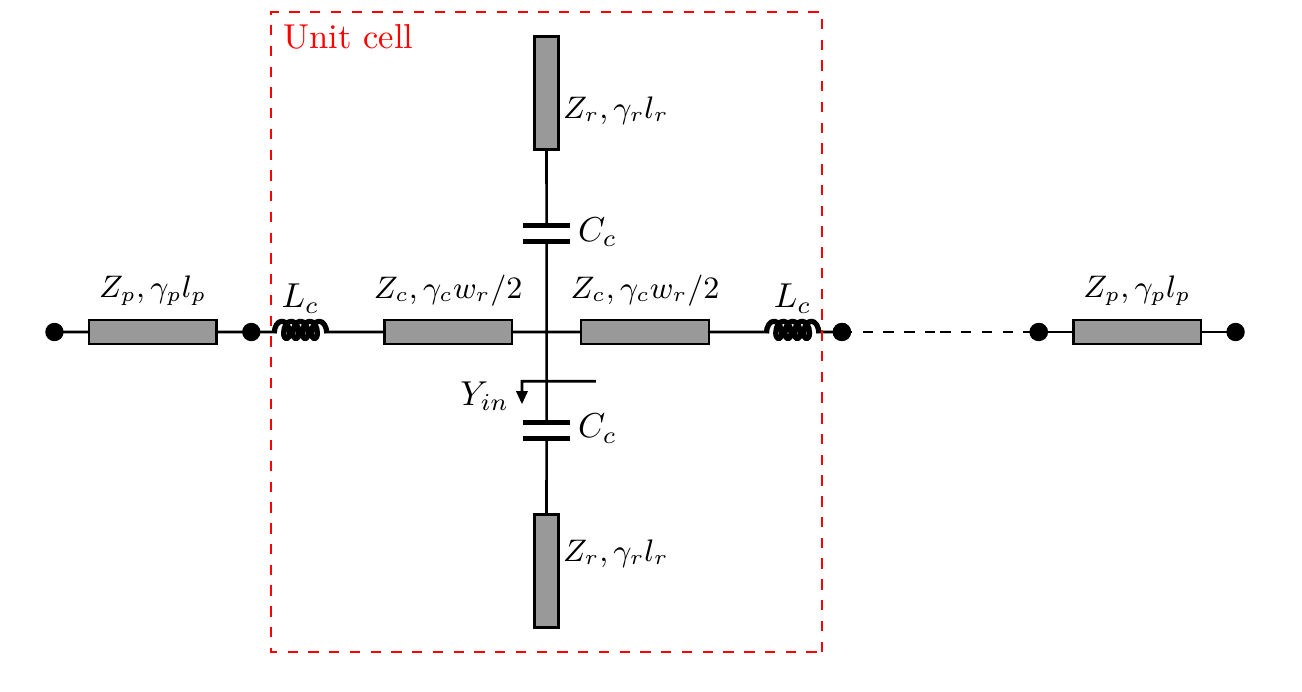}
\\
\caption{Transmission-line model of the periodic filter consisting of repeating unit cells of capacitively-coupled open-ended resonators.}
\label{fig:cell_model}
\end{figure}

The proposed terahertz filter consists of a PGL periodically loaded with capacitively-coupled $\lambda/2$ resonators in a balanced configuration, see Fig.\,\ref{fig:filter}.
The resonators are physically separated from the PGL to not short-circuit its propagating mode but close enough to produce coupling between the PGL and resonators.
The input impedance presented by the shunt resonators along the line creates a frequency response with stopbands \cite{Young1966} when the resonators exhibit impedances close to a short-circuit.
The periodic structure of $N$ equal resonator pairs in cascade increases the bandwidth of the stopbands and, thereby, produces sharper and narrower passbands.
The layout parameters of the filter (Fig.\,\ref{fig:filter_pic_zoom}) are the PGL's width, $w_p$; resonator's length, $l_r$, gap, $g_r$, and width, $w_r$; and the line-resonator coupling gap, $g_c$.

This section describes the transmission-line model, design, fabrication process, and on-wafer measurement setup to demonstrate a bandpass PGL filter centered at 0.9\,THz fabricated on a silicon membrane.

\subsection{Transmission-line model} \label{sec:model}

Herein, an equivalent transmission-line model \cite{Marks1992} is proposed to explain the rather complex electromagnetic wave propagation through the filtering structure.
The filter, consisting of repeating resonators along the PGL, can be modeled as periodic unit cells of length $l_c=g_r+w_r$.
Each unit cell represents a section of the PGL central filter strip with a coupled resonator-pair and two half resonator-gaps on both sides, modeled as shown in Fig.\,\ref{fig:cell_model}.
The inductances, $L_c$, represent an electrically-short high-impedance line section with length $g_r/2$.
\textcolor{black}{The unit-cell model does not consider any coupling to neighboring resonators.}

Two types of planar waveguides are used in the unit-cell model: coplanar waveguides (even mode) for the central line crossing the filter, $\gamma_c$ and $Z_c$; and parallel metal strips (even mode) for the waves excited in the coupled resonators, $\gamma_r$, and $Z_r$.
No coplanar waveguide odd modes will be excited in the central line due to the symmetry of the filter \cite{Ribo2000}.
Since the separation of contiguous resonators is much smaller than the wavelength, the excitation of the odd coplanar-stripline mode can be neglected \cite{Ponchak2018}.
Higher-order modes in the PGL decay rapidly, similarly to the Goubau line's \cite{Zelby1962,Fikioris1979}, and therefore were also neglected.
Naturally, the line feeding the filter was modeled as a PGL with $\gamma_p$ and $Z_p$.

The resonator's electrical length, $\gamma_r l_r$, determines the resonance frequency of the filter's stopband, as it can produce an input admittance, $Y_{in}$, in shunt with the main line, which tends to infinity at the stopband frequencies.
The resonant frequencies of the capacitance-resonator can be found by applying transmission-line theory and calculating $Y_{in}$ as a function of frequency.
$Y_{in}$ for a single coupled resonator is:
\begin{equation}
\label{eq:Yfilter}
    Y_{in} = \frac{j\omega C_c \tanh{\gamma_r l_r}}{\tanh{\gamma_r l_r} + j\omega C_c Z_r},
\end{equation} 
where $\omega$ is the angular frequency and $C_c$ is the coupling capacitance.
The properties of the unit cell shown in Fig.\,\ref{fig:cell_model} can be calculated by cascading the \textit{ABCD} matrices of each element as:
 
\begin{equation}
\begin{split}
[U] = [C][L][Y][L][C],
\end{split}
\end{equation}
where $[C]$, $[L]$, and $[Y]$ are the $ABCD$ matrices of a coplanar waveguide, a series inductor, and a shunt  $Y_{in}$ admittance, given by:
 
\begin{subequations}\label{eq:5}
\begin{align}
 [C] &= 
 \begin{bmatrix}
 \cosh{\left(\gamma_c w_r/2\right)} & \sinh{\left(\gamma_c w_r/2\right)} Z_c  \\
 \sinh{\left(\gamma_c w_r/2\right)}/ Z_c & \cosh{\left(\gamma_c w_r/2\right)}
 \end{bmatrix}\\
 [L] &= 
 \begin{bmatrix}
 1 & j\omega L_c  \\
 0 & 1
 \end{bmatrix}\\
 [Y] &=
 \begin{bmatrix}
 1 & 0  \\
 Y_{in} & 1
 \end{bmatrix}.
\end{align}
\end{subequations}
 
Then, the \textit{ABCD} matrix of the proposed model with a filter with $N$ unit cells and a total PGL length of $l_t$ is

\begin{equation} \label{eq:filterMatrix}
 {
 [F] = 
 [P]
 [U]^{N}
 [P]
 },
 \end{equation}

 where $[P]$ is the \textit{ABCD} matrix corresponding to a PGL with length $l_p=(l_t-Nl_c)/2$ (see Fig.\,\ref{fig:cell_model})

 \begin{equation} \label{eq:PGLMatrix}
 {
 [P]=
 \begin{bmatrix}
 \cosh{\left(\gamma_p l_p\right)} &  \sinh{\left(\gamma_p l_p\right)}Z_p \\
 \sinh{\left(\gamma_p l_p\right)}/Z_p  & \cosh{\left(\gamma_p l_p\right)} 
 \end{bmatrix}
 }.
 \end{equation}

 \subsection{Filter design} \label{sec:design}
 
A 0.9-THz passband filter was designed, \textcolor{black}{simulated} and fabricated on a $10\text{-}\mu\text{m}$ thick high-resistivity ($\rho > \SI{10}{k\ohm.\cm}$) suspended silicon membrane, with a bulk permittivity of $\epsilon_r=11.7$ and $\tan\delta=1.7\cdot10^{-5}$ \cite{Dai2004}, \textcolor{black}{and using gold as conductor ($\sigma=\SI{4.1e7}{S/m}$)}.
An electrically-thin suspended substrate helps reduce dielectric and radiation losses, crucial at THz frequencies.
The radiation losses are minimized by having a higher phase constant, $\beta$, in the planar waveguides than in the undesired substrate modes \cite{Rutledge1978,Rutledge1983_thinCPW,Grischkowsky1987}.
With the proposed substrate, the critical frequencies for coupling to substrate modes \cite{Hatkin1954} are calculated to be beyond 5\,THz, and for second-order substrate modes, the cut-off frequency is around 4.6\,THz.

The PGL was designed with a relatively large strip width of $w_p=10\, \mu\text{m}$, which decreases conductor loss \cite{Xu2006b}.
For the resonator width and gap, we took $w_r=5\,\mu\text{m}$ and $g_r=5\, \mu\text{m}$ to guarantee sub-wavelength structure and operate well below the \textcolor{black}{Bragg} frequency \textcolor{black}{\cite{Harvey1960}}.
In our case $\text{Im}(\gamma_c) (w_r+g_r) = \pi$ around 5\,THz, thus we operate well below the \textcolor{black}{Bragg} frequency.
We chose a coupling gap of $g_c=10\,\mu\text{m}$, which coupled the resonators while not short-circuiting the line's propagating mode.

\begin{figure}
\subfloat[]{%
\centering%
\centering
\includegraphics[width=0.49\textwidth]{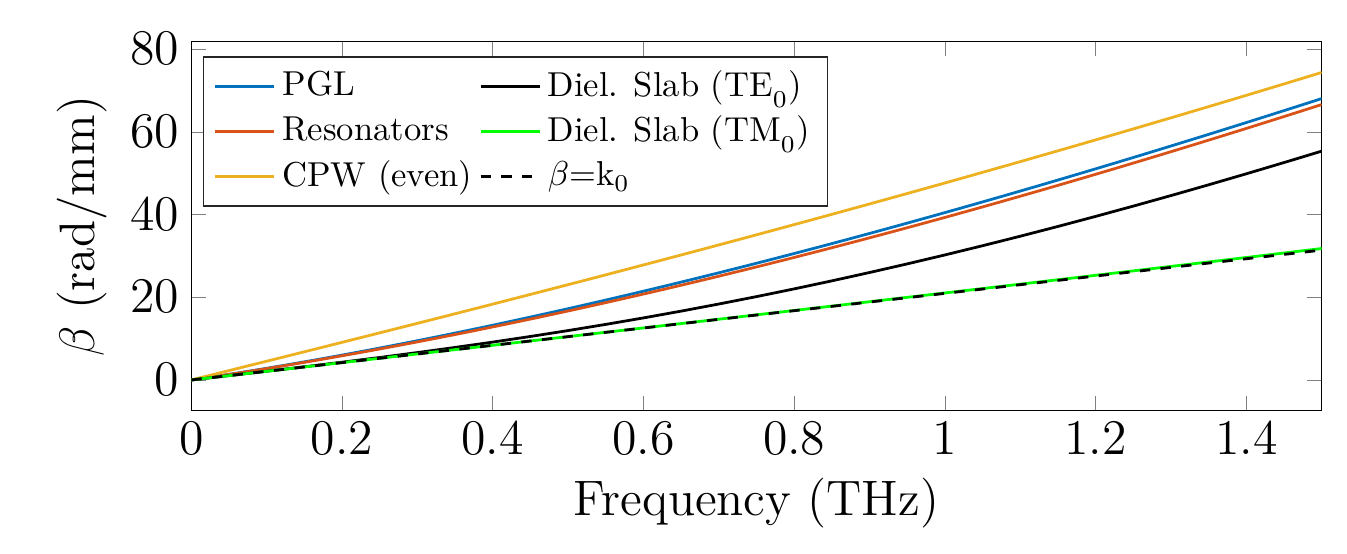}
\label{fig:betaPGL}
}
\\[-2mm]%
\subfloat[]{%
\centering
\includegraphics[width=0.49\textwidth]{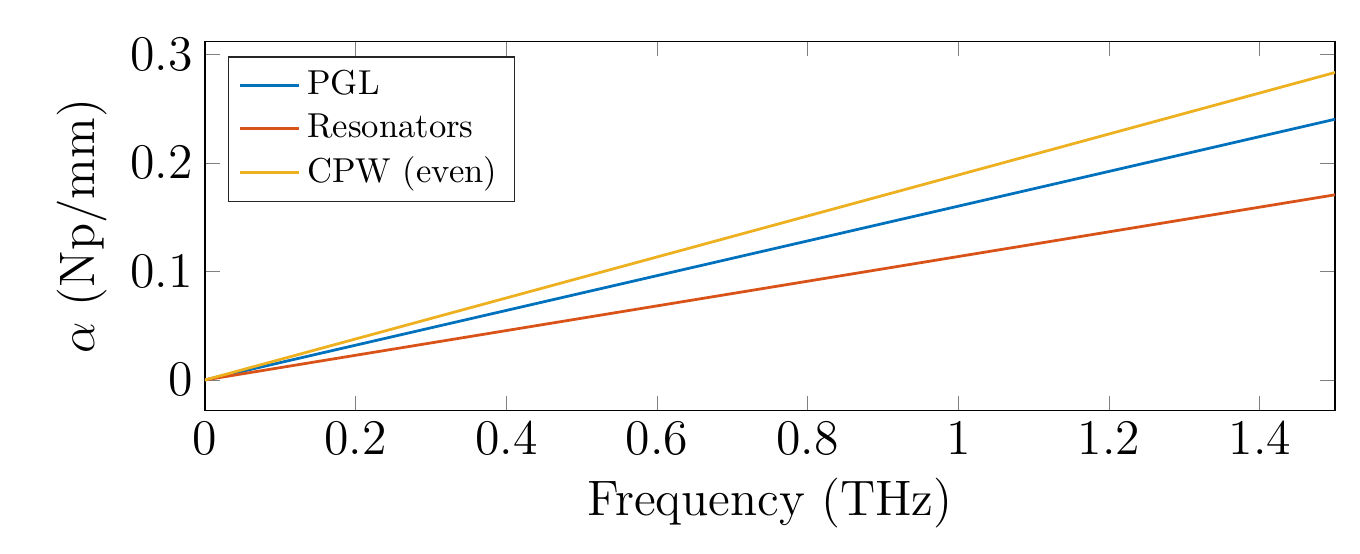}
\label{fig:alphaPGL}
}%
\\[-2mm]%
\subfloat[]{%
\centering
\includegraphics[width=0.49\textwidth]{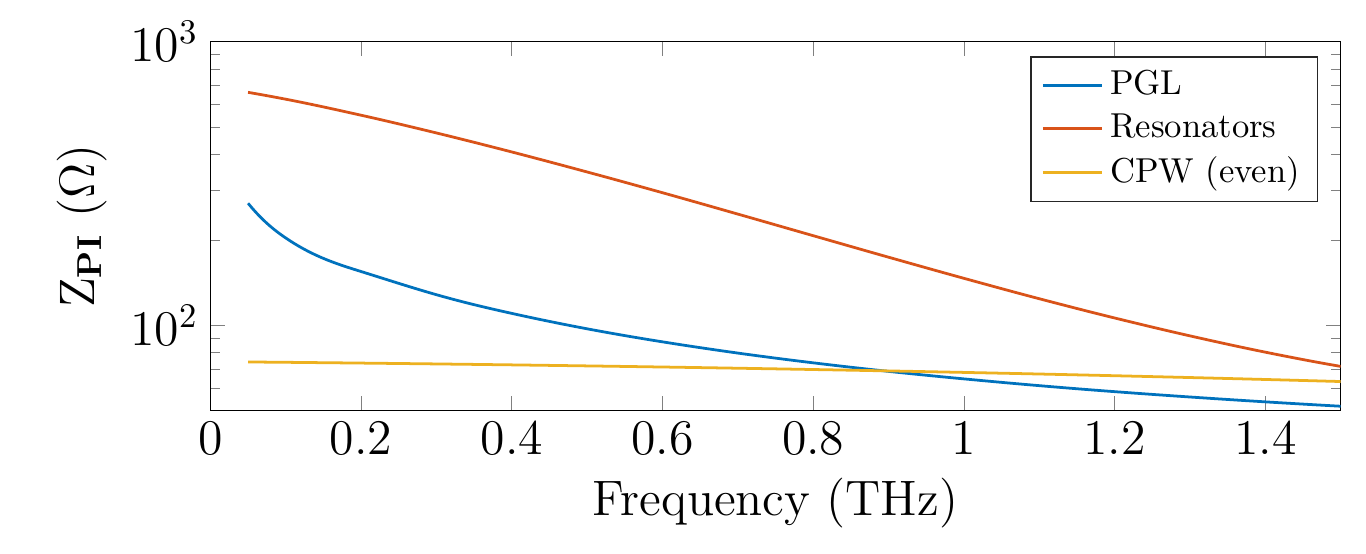}
\label{fig:ZPGL}
}%
\\
\caption{Simulated wave-propagating properties vs. frequency for PGL, resonators, and coplanar waveguide (CPW), for the used substrate. \textbf{(a)} Phase constant of the planar waveguides compared to the dielectric-slab modes. \textbf{(b)} Attenuation constants. \textbf{(c)} Characteristic impedance defined from eq.\,(\ref{eq:Z}).}
\label{fig:alphaBetaZPGL}
\end{figure}

After choosing the substrate and filter layout dimensions, the propagation constants and impedances of the three lines involved in the model were calculated with electromagnetic simulations using absorbing boundary conditions.
Port sizes were a vacuum wavelength at the lowest frequency of analysis for PGL and resonators, and  $3(w_p+2g_c) \text{ by } 2(w_p+2g_c)$ for the rectangular port for the coplanar waveguide.

Each planar waveguide's propagation constant, $\gamma = \alpha + j\beta$, was obtained from the S-parameter results using 3D electromagnetic simulations based on the finite integration technique (CST Studio Suite).
Simulation results were calculated between 50\,GHz and 1.5\,THz in several frequency bands and extrapolated down to dc.
The excitation ports used inhomogeneous port accuracy enhancement, enabling accurate wideband results.
The phase constants of the dielectric slab modes were calculated analytically  \cite{Hatkin1954}.
The dispersion diagrams of the simulated planar waveguides used to model the filter are shown in Fig.\,\ref{fig:betaPGL} together with the fundamental transverse-electric and transverse-magnetic modes of the dielectric slab mode.
As desired, all the planar waveguides show a higher phase constant than the dielectric slab modes present in the substrate, which minimizes radiation loss.
The attenuation constants of the planar waveguides, $\alpha$, are shown in Fig.\,\ref{fig:alphaPGL}, presenting results proportional to frequency.



The characteristic impedance of the lines was calculated using 2D finite-element-method simulations (COMSOL multiphysics) of the cross-section of the planar waveguides between 50\,GHz and 1.5\,THz, using the power-current definition \cite{Schelkunoff1944}:
\begin{equation} \label{eq:Z}
    Z_{PI} = \frac{2P_{avg}}{\lvert I \rvert^2} = 2\frac{\iint_A \langle \Vec{S} \rangle \,d\Vec{A}}{\left\lvert \oint_C \Vec{H} \,d\Vec{l} \right\rvert^2}
\end{equation}

where $P_{avg}$ denotes the time-averaged power of the waveguide, and $I$ is the current.
The characteristic impedance results of the planar waveguides used for modeling the filter are shown in Fig.\,\ref{fig:ZPGL}, \textcolor{black}{neglecting the imaginary part of the PGL's impedance \cite{Cavallo2017}}.

\begin{figure}
\subfloat[]{%
\centering%
\centering
\includegraphics[width=0.49\textwidth]{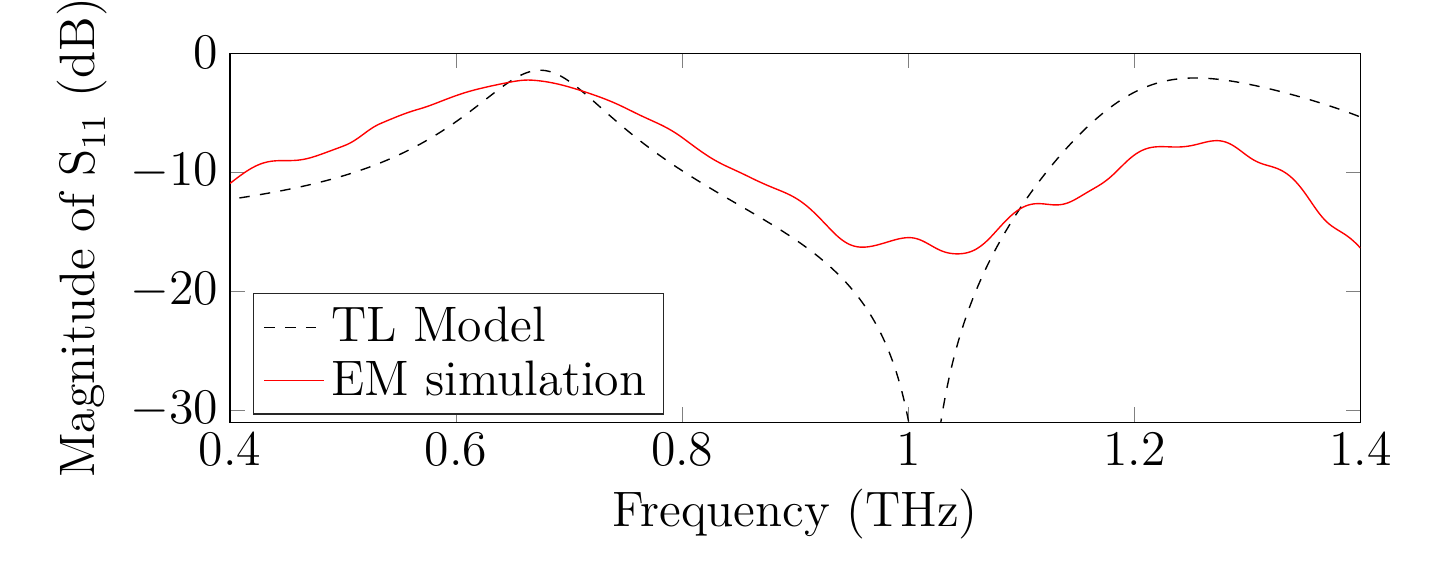}
\label{fig:ModSim_S11_1cell}
}%
\\[-2mm]%
\subfloat[]{%
\centering
\includegraphics[width=0.49\textwidth]{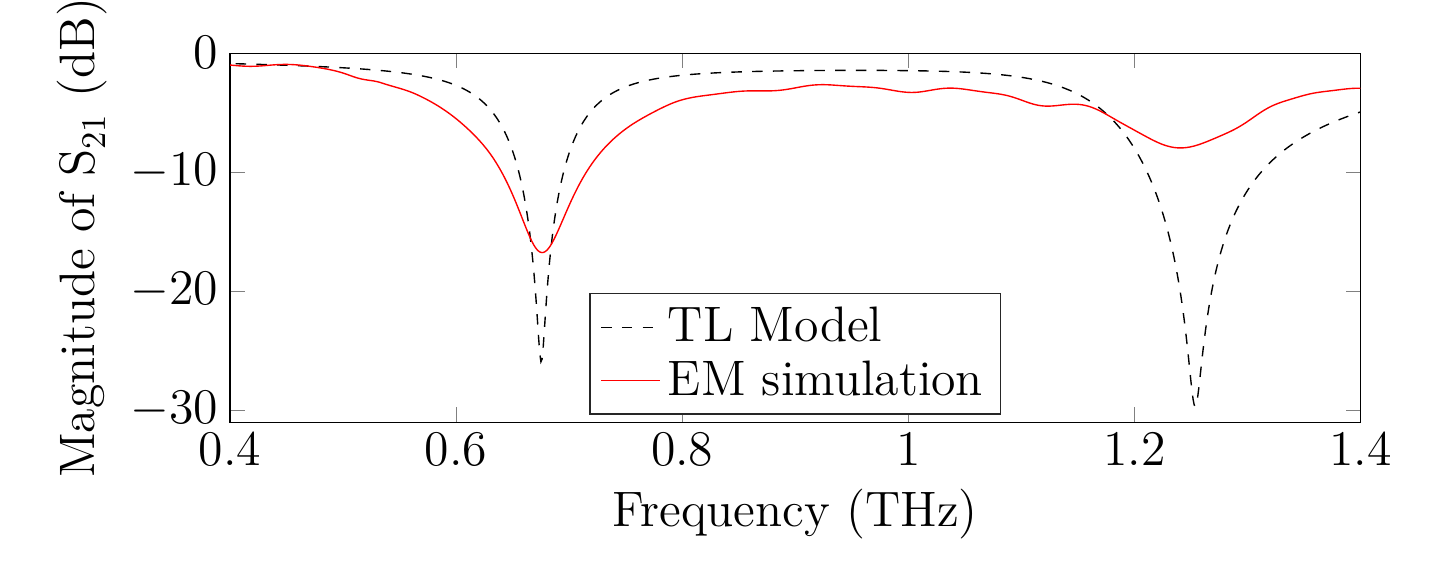}
\label{fig:ModSim_S21_1cell}
}%
\\
\caption{S-parameter comparison of the transmission-line model and electromagnetic simulations for a single-resonator-pair filter. The model shows to properly reproduce the pass- and stopbands of the filter. (a)  $\lvert S_{11} \rvert$ and (b) $\lvert S_{21} \rvert$.}
\label{fig:ModSim_1cell}
\end{figure}

With the above values of $w_r$, $g_r$ and $g_c$, we fitted the values of $C_c$ and $L_c$ to 1.5\,fF and 7\,pH, respectively, by comparing 3D electromagnetic simulations of a single resonator pair of arbitrary length with its model, taking $\gamma_r$ and $Z_r$ from a $5\text{-}\mu\text{m}$ PGL strip.
With the resulting values of $C_c$ and $L_c$, a stopband centered around 0.6\,THz requires a resonator length of $l_r=100\,\mu\text{m}$, producing a passband response around 0.9\,THz.
Fig.\,\ref{fig:ModSim_1cell} compares the scattering parameters of the proposed model and electromagnetic simulation results of a filter with a single resonator pair of length $l_r=100\,\mu\text{m}$, showing fair agreement, and replicating the resonances in transmission.

\begin{figure}
\centering
\includegraphics[width=0.5\textwidth]{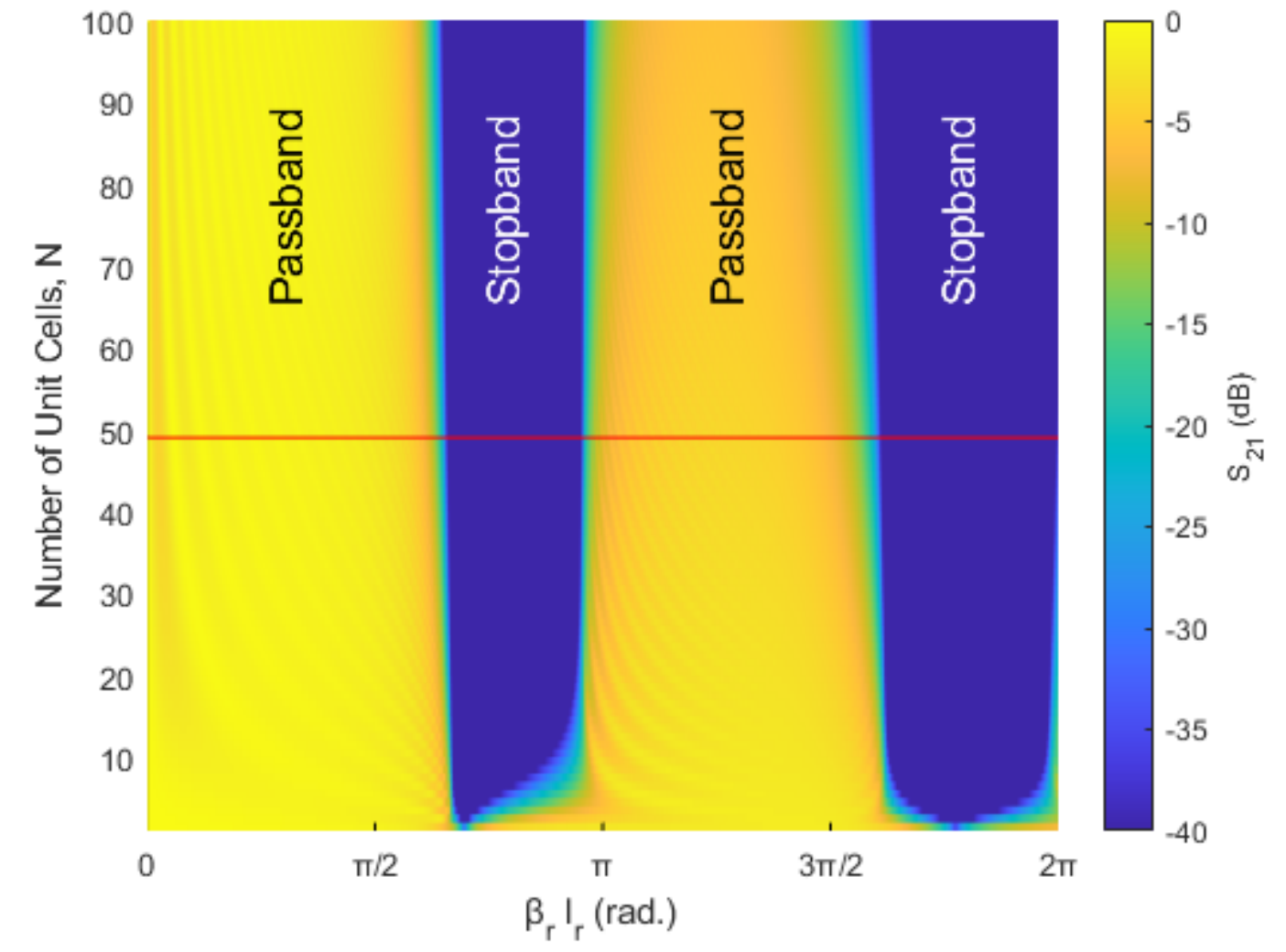}
\\
\caption{Magnitude of transmission, $\lvert S_{21} \rvert$, calculated using the transmission-line model from Fig.\,\ref{fig:cell_model}, of the proposed PGL filter as a function of the number of resonators, $N$, and the normalized frequency, $\beta_r l_r$. Values in dark blue color have a $\lvert S_{21} \rvert \le 40\,dB$.  The fabricated filter has 49 unit cells, marked with the red line.}
\label{fig:Mod_S21_vs_nStubs}
\end{figure}

Fig.\,\ref{fig:Mod_S21_vs_nStubs} shows the modeled transmission, $S_{21}$, of the proposed filter with $l_r=100\,\mu\text{m}$ as a function of the number of unit cells, while maintaining the total length of the PGL at a constant total size of $l_t=1\,\text{mm}$.
As can be seen, the bandwidth of the stopband increases with the number of unit cells and saturates around 20, being confined to regions where $\pi(n-0.5)<\beta l_r<\pi n, \quad \forall n \in \mathbb{N}$.
Due to the capacitive coupling, resonances are up-shifted from odd multiples of $\beta l_r = \pi/2$.
We chose $N=49$ unit cells for the fabricated filter.

Finally, a transition from coplanar waveguide to PGL was included in the layout  (Fig.\,\ref{fig:filter_pic}) for on-wafer characterization using ground-signal-ground probes. 
The transition was designed to minimize insertion losses by exponentially changing its line impedance to reduce reflections and by using a quarter-wave transition at 0.9\,THz to lower line losses \cite{Cabello-Sanchez2018}.

\subsection{Fabrication} \label{sec:fab}

\begin{figure}
\subfloat[]{%
\centering%
\includegraphics[width=0.49\textwidth]{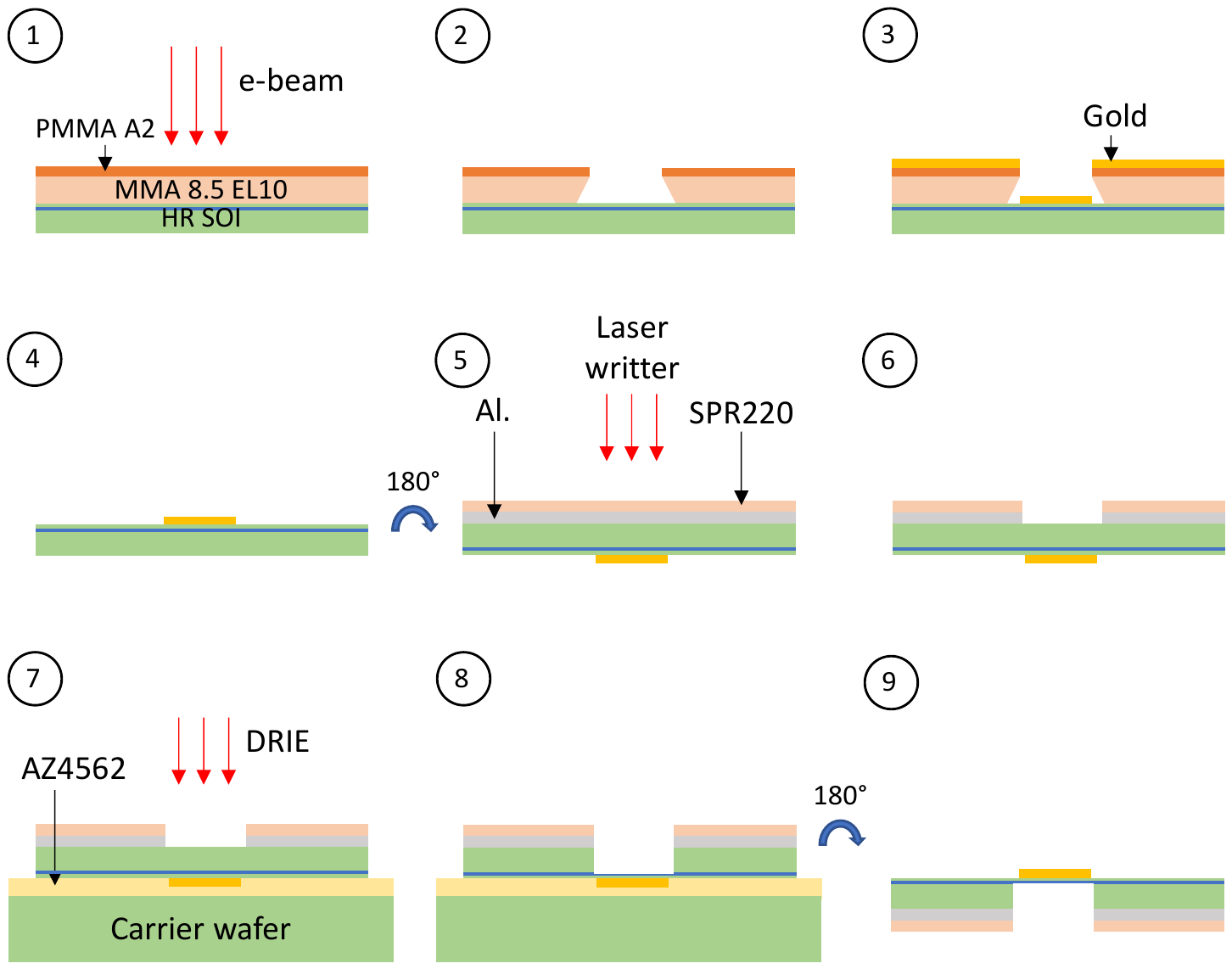}
\label{fig:fabSteps}
}%
\\
\subfloat[]{%
\centering%
\includegraphics[width=0.49\textwidth]{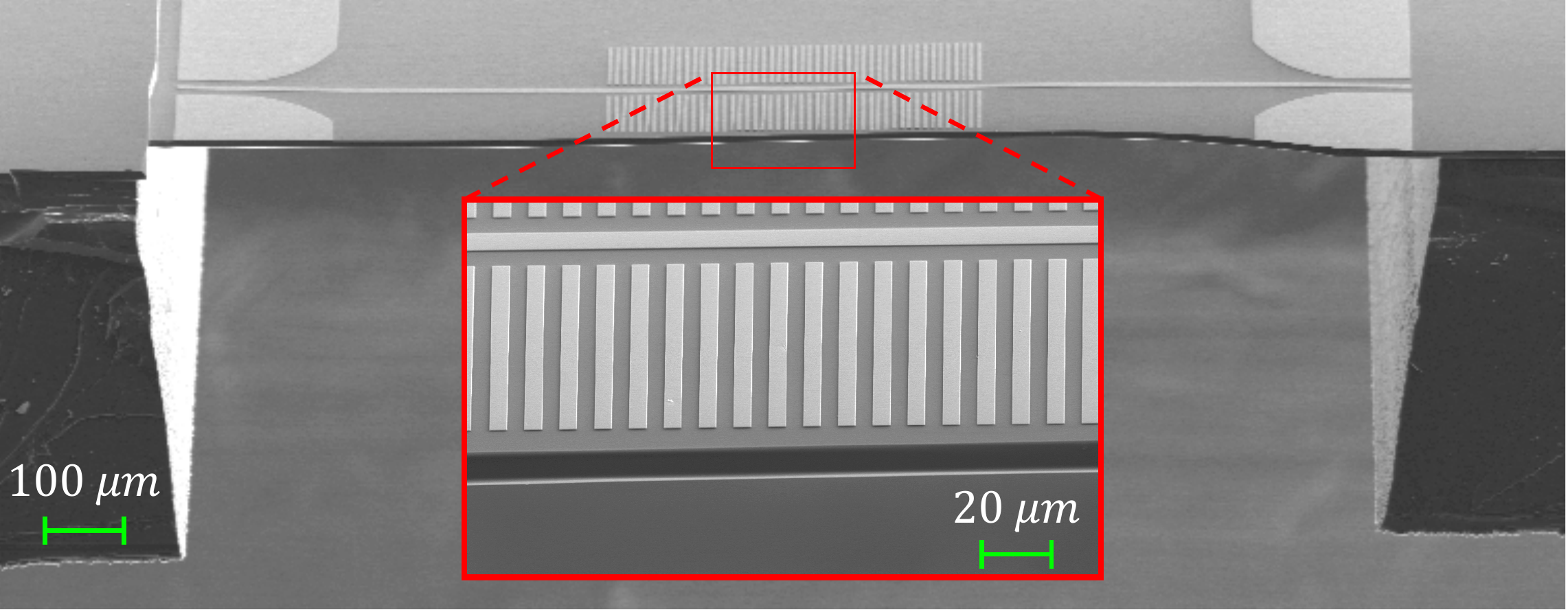}
\label{fig:SEM_pic}
}%
\\[2.6mm]
\caption{(a) Sketch of the fabrication steps of the proposed silicon-on-insulator chip (NB! The drawings are not in scale). (b) Scanning-electron-microscope image of the filter fabricated on the high-resistivity silicon membrane. The inset shows detail of the membrane and resonators. Unfortunately, the left side of the membrane broke when cleaving the chip for the picture.}
\label{fig:fab}
\end{figure}

The suspended PGL filter was fabricated using a silicon-on-insulator (SOI) wafer from University Wafers.
We used a high-resistivity (>$\SI{10}{k\ohm.cm}$), undoped, $10\text{-}\mu\text{m}$ device layer; $1\text{-}\mu\text{m}$ buried oxide layer grown on a high-resistivity, $400\text{-}\mu\text{m}$ thick, silicon handle wafer.

In Fig.\,\ref{fig:fabSteps}, the complete process flow is outlined.
The fabrication process begins by  defining the circuit using electron-beam lithography \textcolor{black}{(\#1,2)}, followed by metal deposition \textcolor{black}{(\#3)} and lift-off \textcolor{black}{(\#4)}.
The conductor metallization consists of a 10-nm titanium adhesive layer and a 350-nm gold layer, evaporated in an ultra-high vacuum chamber.
Prior to the backside processing, the front side, with the circuit layer, is first protected with a thick resist and then mounted topside-down on a 6-inch carrier wafer.
The backside process starts by sputtering a 20-nm aluminum etch mask layer \textcolor{black}{(\#5,6)}.
Openings for the membrane cavities are then defined in a photoresist (\textcolor{black}{SPR220-3.0}) using a direct laser writer.
\textcolor{black}{The chip is mounted on a silicon carrier wafer using a  photoresist (AZ4562) and baked  \textcolor{black}{(\#7)}.}
Then, the silicon carrier wafer with the chip is dry-etched by deep reactive ion etching (Bosch process) in SF$_6$ and O$_2$ atmosphere, using C$_2$F$_8$ as passivation.
The etch rate is about $1\,\mu\text{m}$ per cycle.
The high selectivity toward dry etching between silicon and silicon dioxide ensures that the etching process will stop when the buried oxide is reached \textcolor{black}{(\#8)}.
After etching, the chip is released from the carrier wafer by dissolving the protecting resist in acetone for 48 hours \textcolor{black}{(\#9)}.
Finally, any remaining resist residues are cleaned using an O$_2$-plasma dry etching.

\begin{figure}
\subfloat[]{%
\centering%
\includegraphics[width=0.4\textwidth]{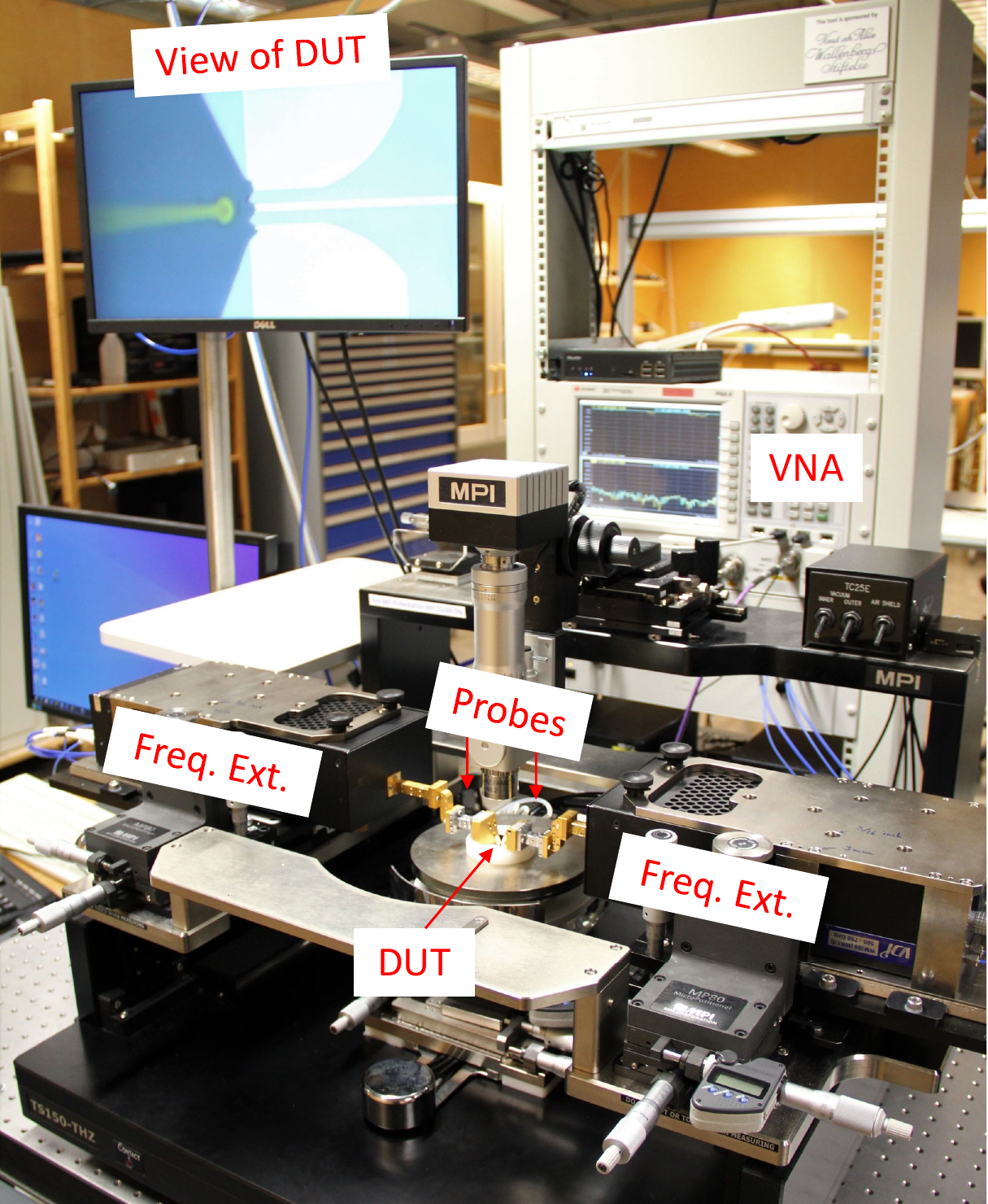}
}%
\\
\centering%
\subfloat[]{%
\centering%
\includegraphics[width=0.35\textwidth]{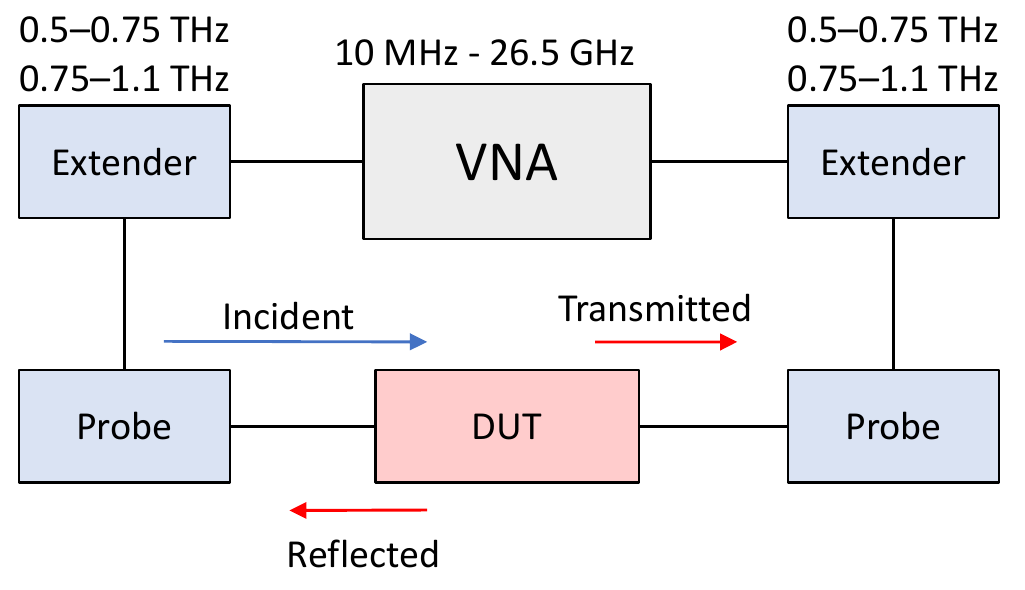}
}%
\\[2.6mm]
\caption{(a) Picture of the measurement setup showing the probe station with the vector network analyzer (VNA), frequency extenders, and on-wafer probes. \textcolor{black}{(b) Sketch of the measurement setup.}}
\label{fig:mSetup}
\end{figure}

A scanning-electron-microscope photograph of the fabricated suspended PGL circuit is shown in Fig.\,\ref{fig:SEM_pic}.
Compared to our previous work using a soft plastic film substrate \cite{Cabello-Sanchez2021}, the suspended silicon substrate has the advantage of having the entire device on top of the suspended membrane, improving the repeatability of measurements and removing supporting substrate interfaces.

\subsection{Measurement set-up} \label{sec:meas}

The PGL filter was characterized by measuring the S-parameters between 0.5\,THz and 1.1\,THz using a vector network analyzer (Keysight N5242A) with frequency extenders (VDI WR1.5SAX and WR1.0SAX) and DMPI T-Wave ground-signal-ground probes \cite{Bauwens2014} (Fig.\,\ref{fig:mSetup}).
The intermediate-frequency bandwidth was set to 100\,Hz.
The measurements were calibrated with dedicated multi-line Thru-Reflect-Line standards for PGL \cite{Cabello-Sanchez2018} fabricated on the same chip.
Calibrating allows to set the reference plane in the PGL (see Fig.\,\ref{fig:filter_pic}) and to de-embed the coplanar waveguide transition.
The lines used for calibration have an electrical length of $\lambda/4$, $3\lambda/4$, and $11\lambda/4$ at 0.91\,THz, with an effective refractive index of $n_e=1.88$.
The chip was placed on top of a polyethylene ($\epsilon_r=2.3$ and $\tan\delta=0.004$ at 1\,THz \cite{Fuse2011}) supporting substrate to isolate the DUT from the probe station's metal chuck.
The suspended $10\text{-}\mu\text{m}$ silicon membrane of the chip has enough mechanical strength to support the pressure from on-wafer probing.

\section{Results} \label{sec:results}

\begin{figure}
\centering
\includegraphics[width=0.49\textwidth]{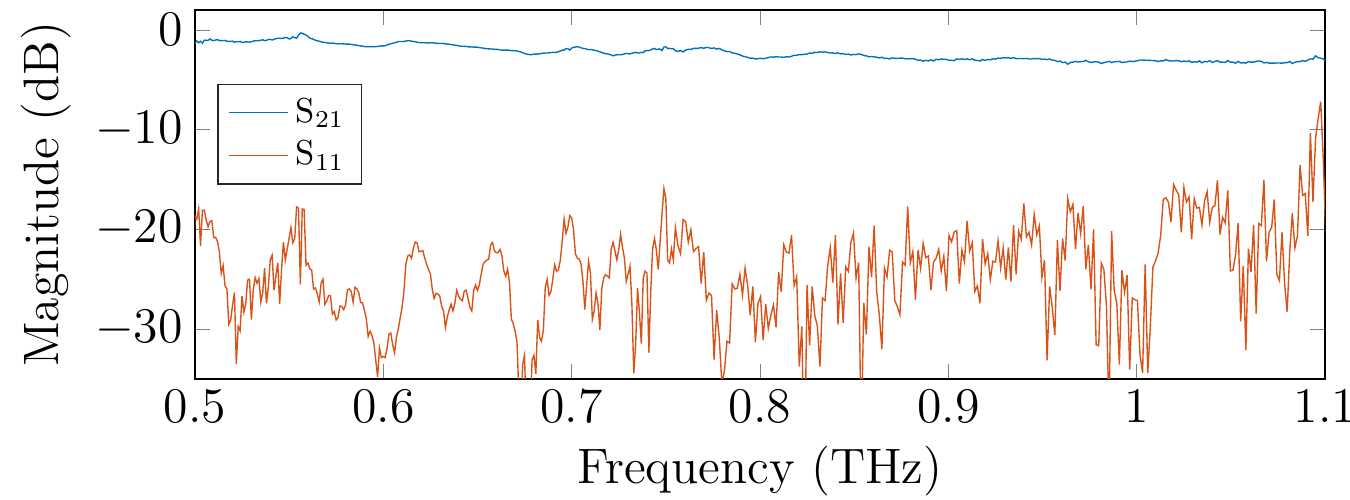}
\caption{Measured  S-parameters for a 1-mm-long PGL fabricated in the same high-resistivity silicon membrane.}
\label{fig:Spar_PGL_1mm}
\end{figure}

\begin{table*}[t]
\centering
\caption{\textcolor{black}{State-of-the-art in planar filters measured above 300\,GHz}}
\label{table1}
\begin{tabularx}{0.7\textwidth}{l c c c c} 
\textbf{$f_c$ (GHz)} & \textbf{3-dB bandwidth (\%)} & \textbf{Waveguide}
& \textbf{Insertion losses}  & \textbf{Reference} \\ 
 \toprule
311 & 7 * & Substrate-integrated waveguide
& \SI{7}{dB} &    \cite{Holloway2020}\\
 \midrule
317 & 2.6 § & Planar Goubau line
& \SI{7.5}{dB} &    \cite{Chen2011}\\
 \midrule
337 & 15 * & Substrate-integrated waveguide
& \SI{2}{dB}  &    \cite{Wang2020}\\
 \midrule
350 & 13 * & Coplanar waveguide 
& \SI{1}{dB} $\dagger$ &    \cite{Ding2021}\\
  \midrule
 354 & 0.13 * & Microstrip 
 & \SI{6}{dB} $\dagger$ &    \cite{PascualLaguna2021}\\
  \midrule
 585 & 24 § 
 & Microstrip & \SI{10}{dB}  &     \cite{Cunningham2005}\\
  \midrule
 \textbf{935} & \textbf{31} * & \textbf{Planar Goubau line} 
 &  $\mathbf{7\,dB}$ &    \textbf{This work} \\
\bottomrule
\end{tabularx}
\caption*{* Passband. § Stopband. $\dagger$ Superconductor
}

\end{table*}

\begin{figure}
\centering
\subfloat[]{%
\centering
\includegraphics[width=0.4\textwidth]{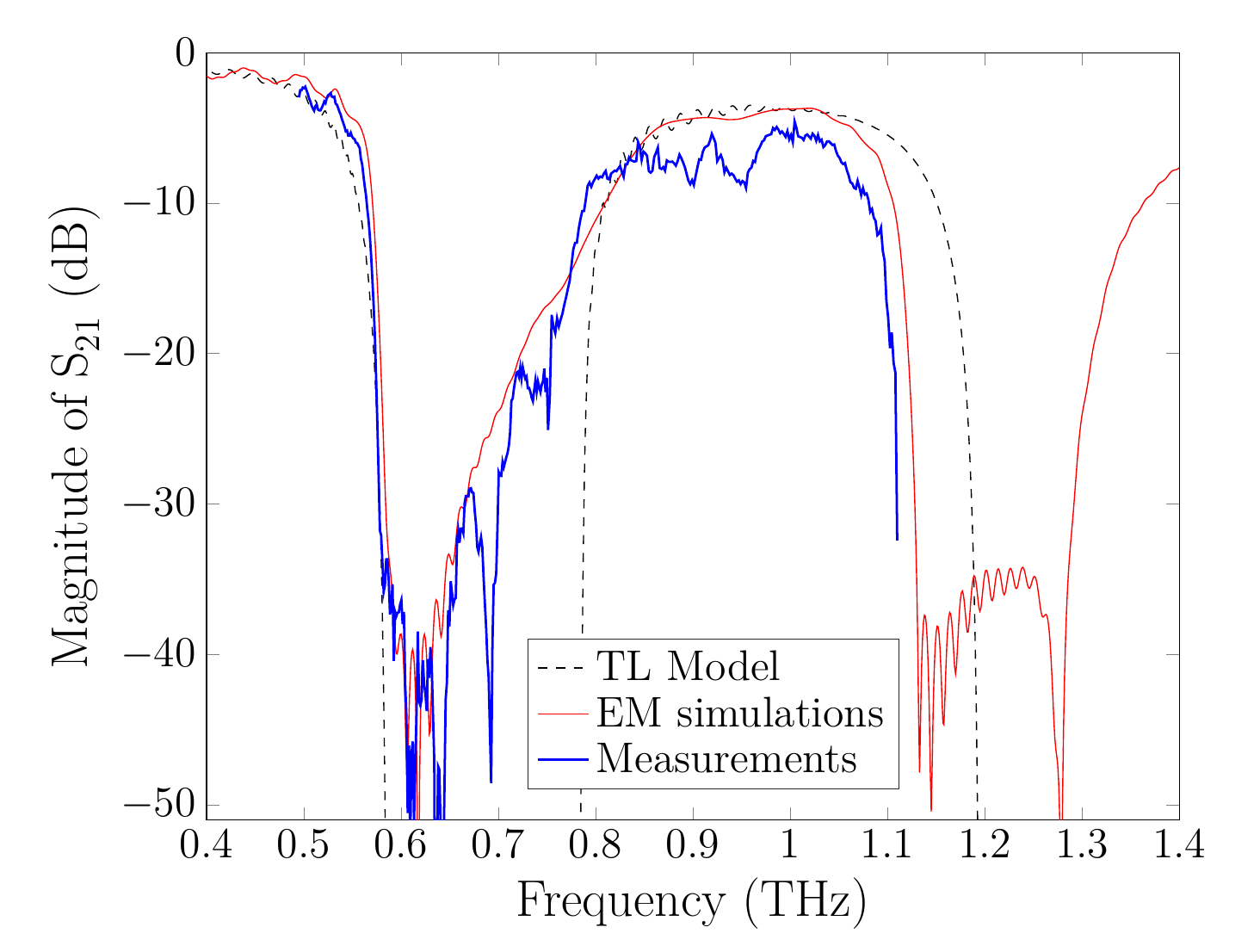}
\label{fig:S21_f100}
}%
\\[-2mm]%
\centering
\subfloat[]{%
\centering
\includegraphics[width=0.4\textwidth]{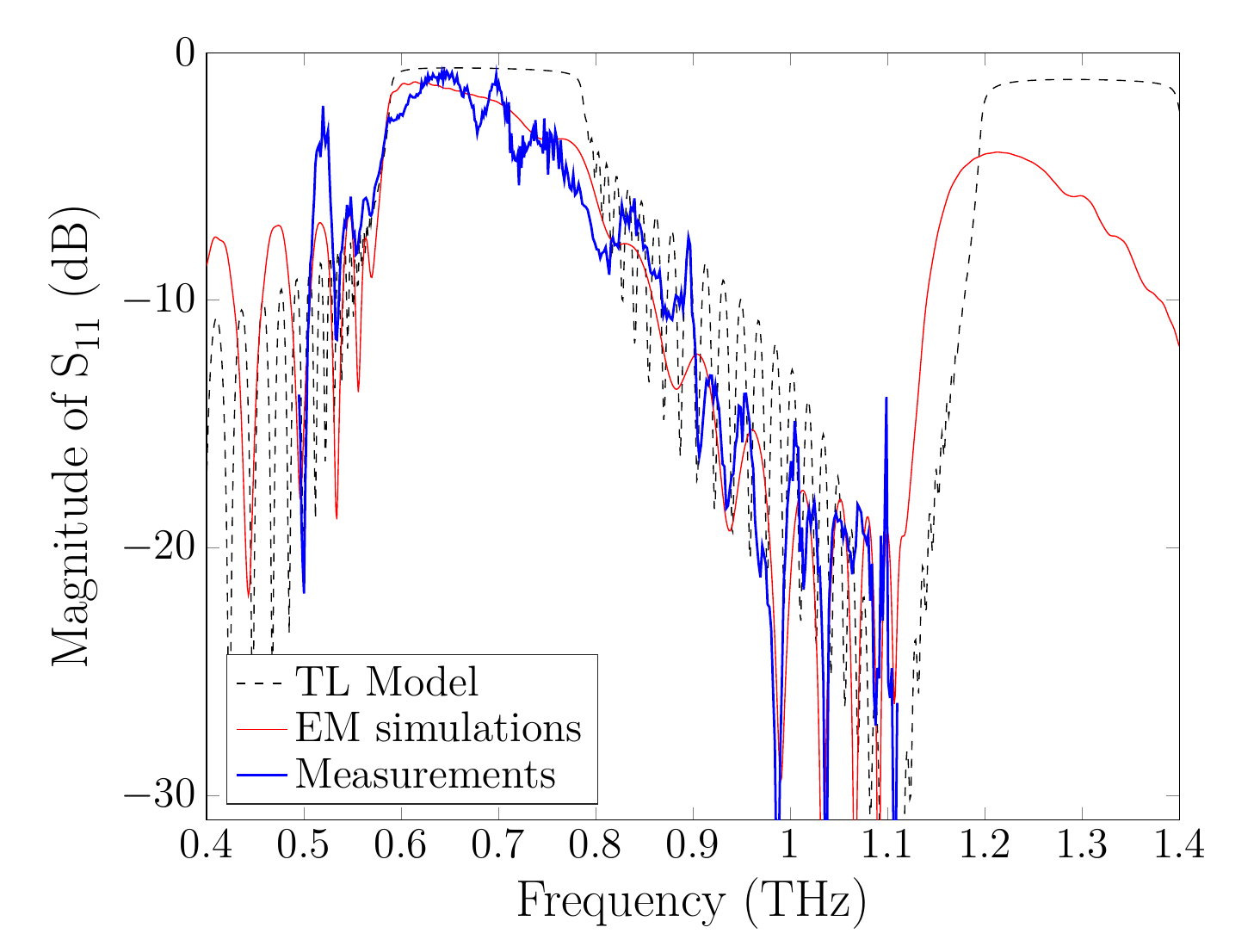}
\label{fig:S11_f100}
}
\\[-0.2mm]%
\centering
\subfloat[]{%
\centering
\includegraphics[width=0.4\textwidth]{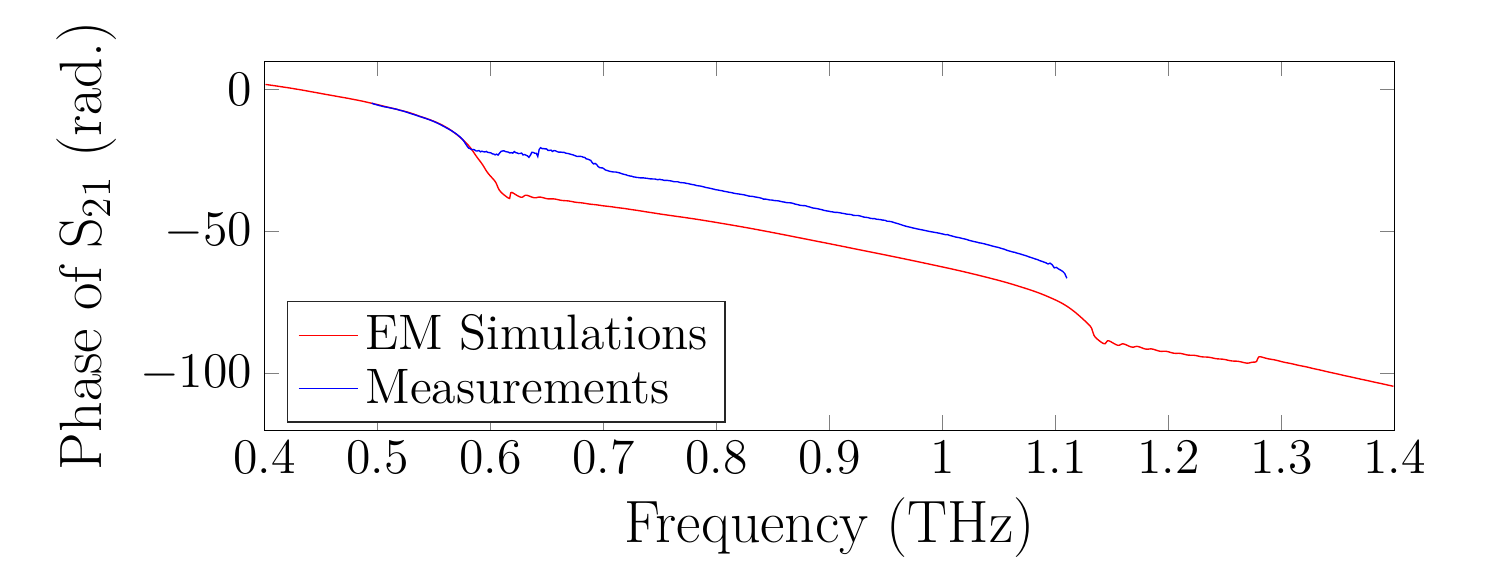}
\label{fig:S21_phase}}
\caption{S-parameter comparison of the transmission-line model, electromagnetic simulations, and measurements for the fabricated 49-unit-cell 0.9\,THz bandpass filter (Fig.\,\ref{fig:filter_pic_zoom}). Measurements agree well with both electromagnetic simulations and the transmission-line model.
(a) shows $\lvert S_{21} \rvert$ results and (b) shows $\lvert S_{11} \rvert$ results, both on decibels.
\textcolor{black}{(c) Phase of $S_{21}$ against frequency, showing low dispersion of the line in the passband.}
}
\label{fig:Spar_filter100}
\end{figure}

\textcolor{black}{
First, Fig.\,\ref{fig:Spar_PGL_1mm} presents the measured S-parameters of a 1-mm-long PGL with no resonators.
The loss goes from 1\,dB/mm at 0.5\,THz to 3\,dB/mm for 1.1\,THz, having relatively low losses for a metal planar waveguide at THz frequencies.
The magnitude of $S_{11}$ is at the noise floor of the measurement setup. 
}


In Fig.\,\ref{fig:Spar_filter100}, the measured S-parameters of the filter from 0.5\,THz to 1.1\,THz are shown together with 3D simulations and the transmission-line model.
The results from electromagnetic simulations show excellent agreement with measurements.
\textcolor{black}{The minor difference between simulation and experimental results in $S_{21}$ can be explained by fabrication tolerances, variation in thin-film material properties, and calibration uncertainties.
The transmission-line model describes well the passband/stopband regime with a slight frequency shift which becomes more noticeable at higher frequencies.}
The filter's $S_{21}$ shows a stopband and a passband centered at 0.6\,THz and 0.9\,THz, respectively, with 3-dB bandwidth of  27\% and 31\%, respectively.
The bandpass has insertion loss of 7\,dB\textcolor{black}{---or about 4\,dB more than a PGL of the same length without resonators (Fig.\,\ref{fig:Spar_PGL_1mm})---which are competitive compared to planar filters measured in the sub-millimeter band (Table\,\ref{table1}).}
Both $S_{11}$ and $S_{21}$ show an asymmetrical passband response, with steeper band transitions at higher passband frequencies.
This effect could be explained by an increase in the effective refractive index at higher passband frequencies caused by the periodic resonators in the filter.
The ripples \textcolor{black}{in $S_{11}$} result from the impedance mismatch between the main PGL and the Bloch impedance of the periodic filter.
\textcolor{black}{The noise level limits in the measurement setup are around -50\,dB for $S_{21}$, and around -22\,dB for $S_{11}$.
Measurements showed a high degree of repeatability across different probe landings, filter structures, and calibrations.}

\textcolor{black}{The measured phase of $S_{21}$ is plotted in Fig.~\ref{fig:S21_phase}, showing low measured dispersion in the passband, and good agreement with simulated results.
The transmission-line model overestimates the phase's slope of the filter, and therefore it was omitted.}

Although simplified, the proposed transmission-line model describes well the magnitude response of the filter demonstrating that the operating principle of the filter can be mostly explained by the periodic placement of coupled $\lambda/2$ resonators and that other possible effects like coupling between neighboring resonators do not play a key role.
The model accuracy could be further improved by recalculating $C_c$ for each of the 49 unit cells---since we kept the same value calculated for a single unit cell---, and taking into account additional parasitic circuit elements.

\begin{figure}
\subfloat[]{%
\centering
\includegraphics[height=0.17\textwidth]{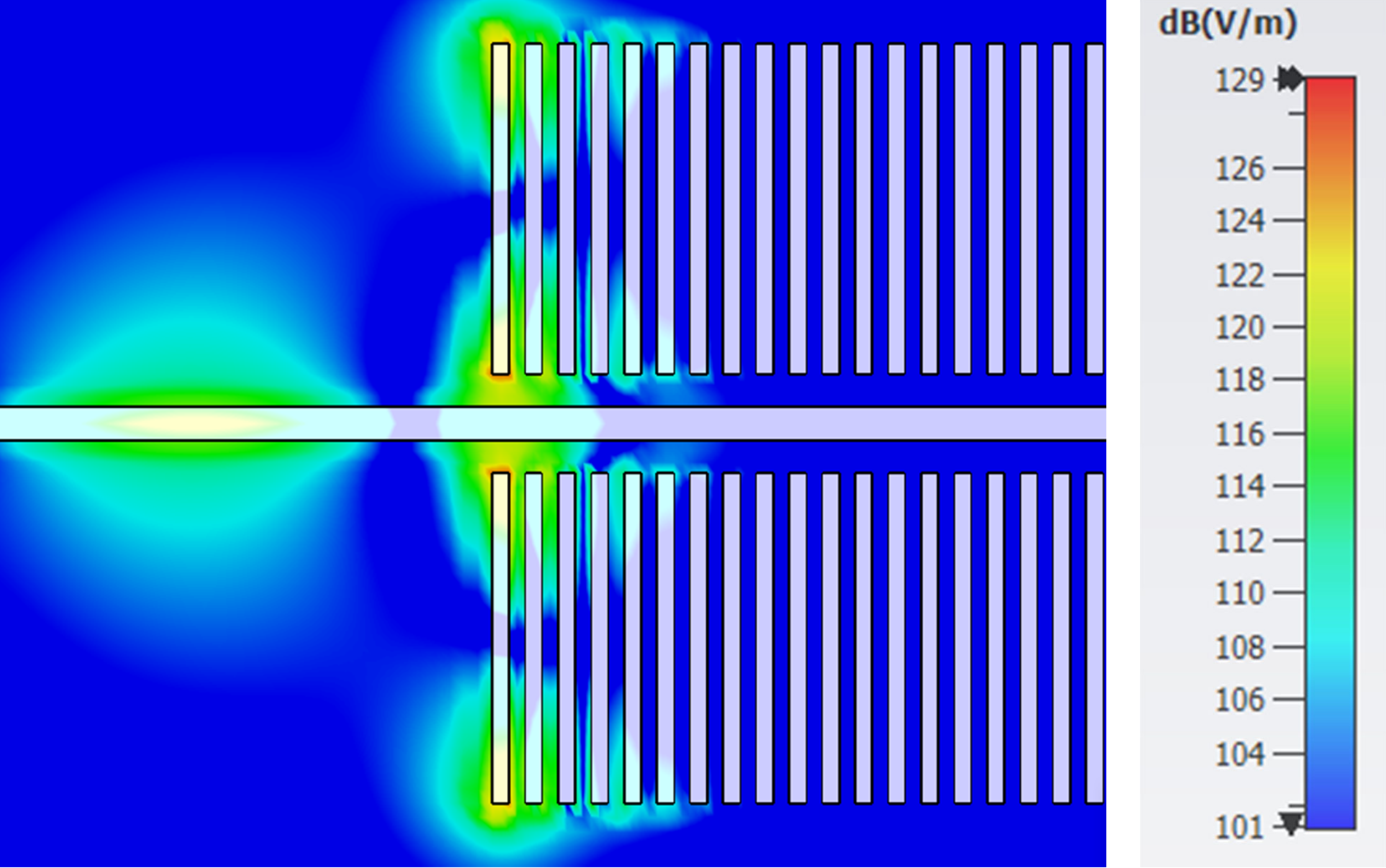}
\label{fig:E600}
}%
\subfloat[]{%
\centering
\includegraphics[height=0.17\textwidth]{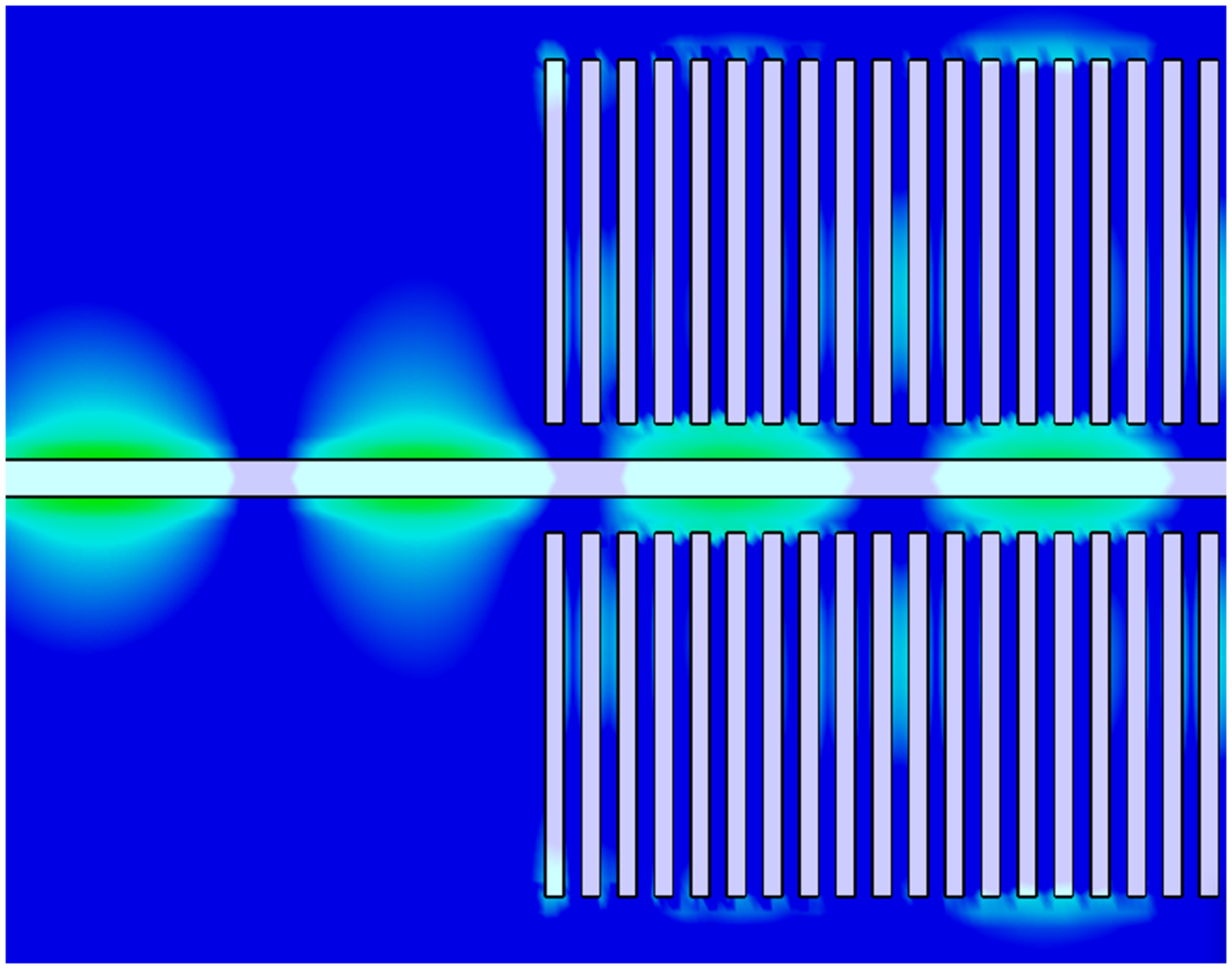}
\label{fig:E900}
}%
\\[2.6mm]
\caption{Simulations of the electromagnetic fields in the PGL filter. The magnitude of the electric field in logarithmic scale at (a) stopband frequency, 0.6\,THz, and at (b) passband frequency, at 0.9\,THz. Scale bar applies to both subfigures.}
\label{fig:Efilt}
\end{figure}

The simulated electric field distribution of the filter in Fig.\,\ref{fig:Efilt} explains the filtering behavior.
When operating at the stopband, at 0.6\,THz (Fig.\,\ref{fig:E600}), the resonators resonate and produce a shunt short-circuit in the main PGL, which prohibits wave propagation.
Fig.\,\ref{fig:E900} shows the E-field of the filter when operating at passband at \textcolor{black}{0.9}\,THz, where the waves can propagate through the filter with a propagating mode similar to that of a coplanar waveguide's even mode.

\begin{figure}
\centering
\includegraphics[width=0.4\textwidth]{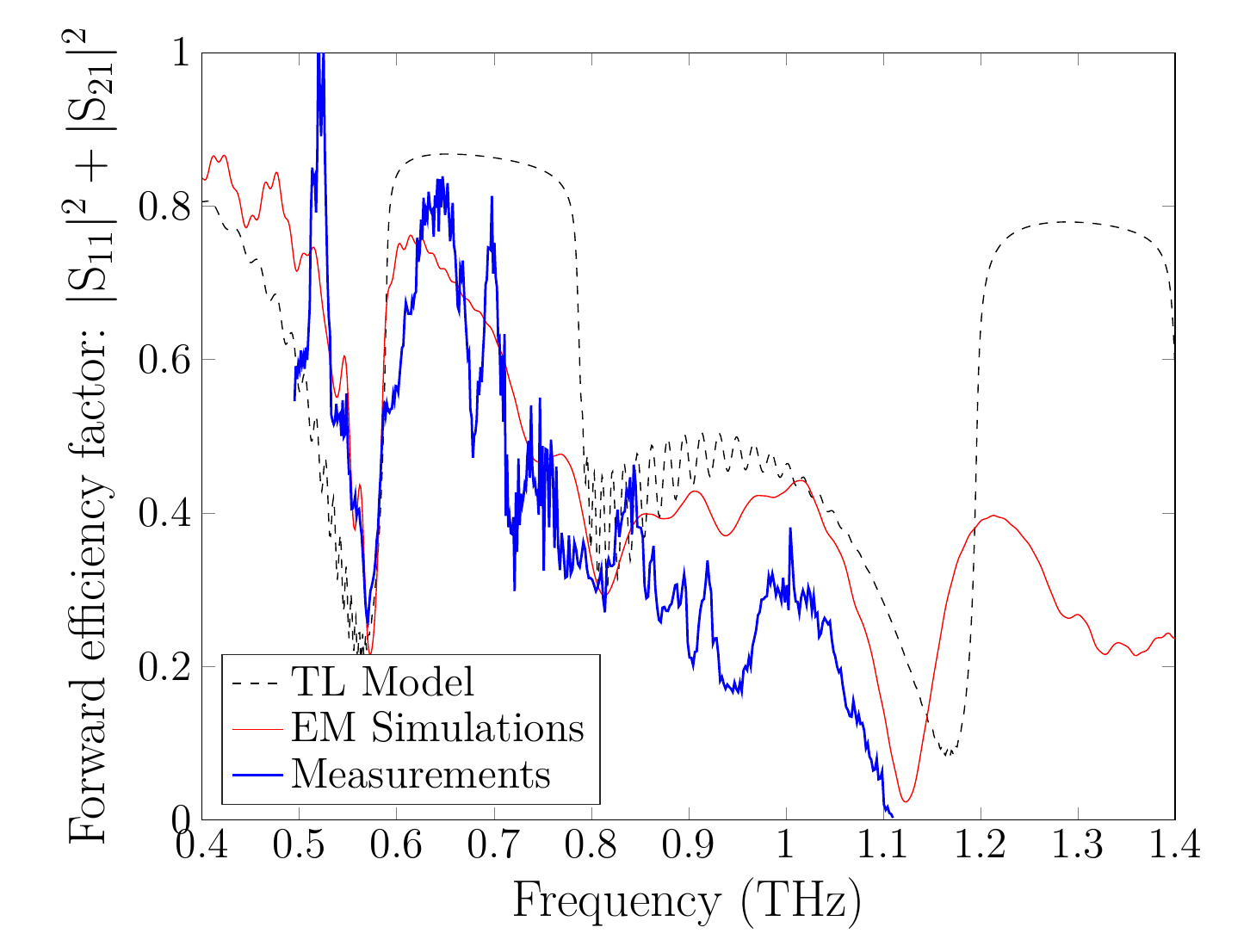}
\caption{Forward efficiency factor vs. frequency, showing the loss of the filter.}
\label{fig:powerBalance_filter}
\end{figure}

The forward efficiency factor ($\lvert S_{11}\rvert^2 +\lvert S_{21}\rvert^2$) is presented in Fig.\, 
\ref{fig:powerBalance_filter}, where increased loss is observed at the transition from a passband to a stopband since currents increase.
\textcolor{black}{
The differences between simulation and measurement results between 0.8 and 1.1\,THz are a consequence of the difference between the levels of $S_{21}$. And similarly, the peak around 520\,GHz can be explained by the spurious peak at the same frequency in $S_{11}$.}

\section{Conclusion} \label{sec:concl}

In summary, we present a planar-Goubau-line terahertz bandpass/bandstop filter \textcolor{black}{which can be easily tuned by changing the electrical length of the periodic capacitively-coupled $\lambda/2$ resonators.}
The working principle of the periodic filter was confirmed by a simple transmission-line model, which agrees with simulations and measurements.
The proposed filter design was validated by fabricating and measuring a periodic structure with 49 unit cells with a passband centered around 0.9\,THz, showing good agreement between simulations and measurements, \textcolor{black}{and a relatively low insertion loss of 7\,dB compared to planar filters measured in the sub-millimeter band.}
Future work could explore the design and tune of a filter response by modulating the length of the different capacitively-coupled resonators \cite{Young1966}, \textcolor{black}{ and further elaborate the transmission-line model}.



%

\appendices

\section*{Acknowledgment}

The authors would like to thank Mr. Mats Myremark for machining parts for the measurement setup; Ms. Divya Jayasankar, Ms. Anis Moradikouchi, and Docent Piotr Starski for their valuable feedback on the manuscript; and the online tool WebPlotDigitizer for its usefulness when extracting data from published graphs.
The devices were fabricated and measured in the Nanofabrication Laboratory and Kollberg Laboratory, respectively, at Chalmers University of Technology, Gothenburg, Sweden.

\ifCLASSOPTIONcaptionsoff
  \newpage
\fi



%


\bibliographystyle{IEEEtran} 
\bibliography{bibtex/bib/IEEEfull.bib, bibtex/bib/references_noURL.bib}

\begin{thebibliography}{10}
\providecommand{\url}[1]{#1}
\csname url@samestyle\endcsname
\providecommand{\newblock}{\relax}
\providecommand{\bibinfo}[2]{#2}
\providecommand{\BIBentrySTDinterwordspacing}{\spaceskip=0pt\relax}
\providecommand{\BIBentryALTinterwordstretchfactor}{4}
\providecommand{\BIBentryALTinterwordspacing}{\spaceskip=\fontdimen2\font plus
\BIBentryALTinterwordstretchfactor\fontdimen3\font minus
  \fontdimen4\font\relax}
\providecommand{\BIBforeignlanguage}[2]{{%
\expandafter\ifx\csname l@#1\endcsname\relax
\typeout{** WARNING: IEEEtran.bst: No hyphenation pattern has been}%
\typeout{** loaded for the language `#1'. Using the pattern for}%
\typeout{** the default language instead.}%
\else
\language=\csname l@#1\endcsname
\fi
#2}}
\providecommand{\BIBdecl}{\relax}
\BIBdecl

\bibitem{Mittleman2017}
D.~M. Mittleman, ``{Perspective: Terahertz science and technology},''
  \emph{Journal of Applied Physics}, vol. 122, no.~23, pp. 1--12, 2017, doi:
  \href{http://dx.doi.org/10.1063/1.5007683}{10.1063/1.5007683}.

\bibitem{Phillips1992}
T.~Phillips and J.~Keene, ``{Submillimeter astronomy (heterodyne
  spectroscopy)},'' \emph{Proceedings of the IEEE}, vol.~80, no.~11, pp.
  1662--1678, 1992, doi:
  \href{http://dx.doi.org/10.1109/5.175248}{10.1109/5.175248}.

\bibitem{Appleby2007a}
R.~Appleby and H.~B. Wallace, ``{Standoff detection of weapons and contraband
  in the 100 GHz to 1 THz region},'' \emph{IEEE Transactions on Antennas and
  Propagation}, vol.~55, no. 11 I, pp. 2944--2956, 2007, doi:
  \href{http://dx.doi.org/10.1109/TAP.2007.908543}{10.1109/TAP.2007.908543}.

\bibitem{Pickwell2006}
E.~Pickwell and V.~P. Wallace, ``{Biomedical applications of terahertz
  technology},'' \emph{Journal of Physics D: Applied Physics}, vol.~39, no.~17,
  pp. 301--310, 2006, doi:
  \href{http://dx.doi.org/10.1088/0022-3727/39/17/R01}{10.1088/0022-3727/39/17/R01}.

\bibitem{Song2011}
H.~J. Song and T.~Nagatsuma, ``{Present and future of terahertz
  communications},'' \emph{IEEE Transactions on Terahertz Science and
  Technology}, vol.~1, no.~1, pp. 256--263, 2011, doi:
  \href{http://dx.doi.org/10.1109/TTHZ.2011.2159552}{10.1109/TTHZ.2011.2159552}.

\bibitem{Bawuah2021}
P.~Bawuah and J.~A. Zeitler, ``{Advances in terahertz time-domain spectroscopy
  of pharmaceutical solids: A review},'' \emph{TrAC - Trends in Analytical
  Chemistry}, vol. 139, p. 116272, 2021, doi:
  \href{http://dx.doi.org/10.1016/j.trac.2021.116272}{10.1016/j.trac.2021.116272}.

\bibitem{Acbas2014}
G.~Acbas, K.~A. Niessen, E.~H. Snell, and A.~G. Markelz, ``{Optical
  measurements of long-range protein vibrations},'' \emph{Nature
  Communications}, vol.~5, pp. 1--7, 2014, doi:
  \href{http://dx.doi.org/10.1038/ncomms4076}{10.1038/ncomms4076}.

\bibitem{Wen1969}
C.~P. Wen, ``{Coplanar Waveguide: A Surface Strip Transmission Line Suitable
  for Nonreciprocal Gyromagnetic Device Applications},'' \emph{IEEE
  Transactions on Microwave Theory and Techniques}, vol.~17, no.~12, pp.
  1087--1090, 1969, doi:
  \href{http://dx.doi.org/10.1109/TMTT.1969.1127105}{10.1109/TMTT.1969.1127105}.

\bibitem{Knorr1975}
J.~B. Knorr and K.~D. Kuchler, ``{Analysis of Coupled Slots and Coplanar Strips
  on Dielectric Substrate},'' \emph{IEEE Transactions on Microwave Theory and
  Techniques}, vol.~23, no.~7, pp. 541--548, 1975, doi:
  \href{http://dx.doi.org/10.1109/TMTT.1975.1128624}{10.1109/TMTT.1975.1128624}.

\bibitem{Grieg1952}
D.~D. Grieg and H.~F. Engelmann, ``{Microstrip—A New Transmission Technique
  for the Kilomegacycle Range},'' \emph{Proceedings of the IRE}, vol.~40,
  no.~12, pp. 1644--1650, 1952, doi:
  \href{http://dx.doi.org/10.1109/JRPROC.1952.274144}{10.1109/JRPROC.1952.274144}.

\bibitem{Akalin2006}
T.~Akalin, A.~Treizebr{\'{e}}, and B.~Bocquet, ``{Single-wire transmission
  lines at terahertz frequencies},'' \emph{IEEE Transactions on Microwave
  Theory and Techniques}, vol.~54, no.~6, pp. 2762--2767, 2006, doi:
  \href{http://dx.doi.org/10.1109/TMTT.2006.874890}{10.1109/TMTT.2006.874890}.

\bibitem{Wu2021a}
K.~Wu, M.~Bozzi, and N.~J.~G. Fonseca, ``{Substrate Integrated Transmission
  Lines: Review and Applications},'' \emph{IEEE Journal of Microwaves}, vol.~1,
  no.~1, pp. 345--363, 2021, doi:
  \href{http://dx.doi.org/10.1109/jmw.2020.3034379}{10.1109/jmw.2020.3034379}.

\bibitem{Temnov2000}
V.~M. Temnov and O.~S. Orlov, ``{The single-strip line is a new type of the
  integrated circuits for microwaves},'' in \emph{2000 10th International
  Crimean Microwave Conference "Microwave and Telecommunication Technology",
  CriMico 2000}.\hskip 1em plus 0.5em minus 0.4em\relax IEEE, 2000, pp.
  359--360, doi:
  \href{http://dx.doi.org/10.1109/CRMICO.2000.1256140}{10.1109/CRMICO.2000.1256140}.

\bibitem{Hong2004}
W.~Hong and Y.~D. Lin, ``{Single-conductor strip leaky-wave antenna},''
  \emph{IEEE Transactions on Antennas and Propagation}, vol.~52, no.~7, pp.
  1783--1789, 2004, doi:
  \href{http://dx.doi.org/10.1109/TAP.2004.829854}{10.1109/TAP.2004.829854}.

\bibitem{Goubau1950}
G.~Goubau, ``{Surface waves and their application to transmission lines},''
  \emph{Journal of Applied Physics}, vol.~21, no.~11, pp. 1119--1128, 1950,
  doi: \href{http://dx.doi.org/10.1063/1.1699553}{10.1063/1.1699553}.

\bibitem{Goubau1951}
------, ``{Single-Conductor Surface-Wave Transmission Lines},''
  \emph{Proceedings of the IRE}, vol.~39, no.~6, pp. 619--624, 1951, doi:
  \href{http://dx.doi.org/10.1109/JRPROC.1951.233782}{10.1109/JRPROC.1951.233782}.

\bibitem{Barlow1953}
H.~Barlow and A.~Cullen, ``{Surface waves},'' \emph{Proceedings of the IEE -
  Part III: Radio and Communication Engineering}, vol. 100, no.~68, pp.
  329--341, 1953, doi:
  \href{http://dx.doi.org/10.1049/pi-3.1953.0068}{10.1049/pi-3.1953.0068}.

\bibitem{Gacemi2013}
D.~Gacemi, A.~Degiron, M.~Baillergeau, and J.~Mangeney, ``{Identification of
  several propagation regimes for terahertz surface waves guided by planar
  Goubau lines},'' \emph{Applied Physics Letters}, vol. 103, no.~19, p. 1119,
  2013, doi: \href{http://dx.doi.org/10.1063/1.4829744}{10.1063/1.4829744}.

\bibitem{Rotmant1950}
W.~Rotman, ``{A Study of Single-Surface Corrugated Guides},'' \emph{Proceedings
  of the IRE}, vol.~39, no.~8, pp. 952--959, 1951, doi:
  \href{http://dx.doi.org/10.1109/JRPROC.1951.273719}{10.1109/JRPROC.1951.273719}.

\bibitem{Maier2006}
S.~A. Maier, S.~R. Andrews, L.~Mart{\'{i}}n-Moreno, and F.~J.
  Garc{\'{i}}a-Vidal, ``{Terahertz surface plasmon-polariton propagation and
  focusing on periodically corrugated metal wires},'' \emph{Physical Review
  Letters}, vol.~97, no.~17, pp. 1--4, 2006, doi:
  \href{http://dx.doi.org/10.1103/PhysRevLett.97.176805}{10.1103/PhysRevLett.97.176805}.

\bibitem{CabelloSanchez2018}
J.~Cabello-S{\'{a}}nchez, H.~Rodilla, V.~Drakinskiy, and J.~Stake,
  ``{Transmission Loss in Coplanar Waveguide and Planar Goubau Line between
  0.75 THz and 1.1 THz},'' in \emph{2018 43rd International Conference on
  Infrared, Millimeter, and Terahertz Waves (IRMMW-THz)}, Nagoya, 2018, pp.
  1--2, doi:
  \href{http://dx.doi.org/10.1109/IRMMW-THz.2018.8510326}{10.1109/IRMMW-THz.2018.8510326}.

\bibitem{Akalin2009}
T.~Akalin and W.~Padilla, ``{Plasmonic waveguides and metamaterial components
  at terahertz frequencies},'' \emph{APMC 2009 - Asia Pacific Microwave
  Conference 2009}, pp. 2444--2446, 2009, doi:
  \href{http://dx.doi.org/10.1109/APMC.2009.5385481}{10.1109/APMC.2009.5385481}.

\bibitem{Dazhang2009}
L.~Dazhang, J.~Cunningham, M.~B. Byrne, S.~Khanna, C.~D. Wood, A.~D. Burnett,
  S.~M. Ershad, E.~H. Linfield, and A.~G. Davies, ``{On-chip terahertz
  Goubau-line waveguides with integrated photoconductive emitters and
  mode-discriminating detectors},'' \emph{Applied Physics Letters}, vol.~95,
  no.~9, p. 092903, 8 2009, doi:
  \href{http://dx.doi.org/10.1063/1.3216579}{10.1063/1.3216579}.

\bibitem{Treizebre2010}
A.~Treizebre, M.~Hofman, and B.~Bocquet, ``{Terahertz Spiral Planar Goubau Line
  Rejectors for Biological Characterization},'' \emph{Progress In
  Electromagnetics Research M}, vol.~14, no. July, pp. 163--176, 2010, doi:
  \href{http://dx.doi.org/10.2528/PIERM10072110}{10.2528/PIERM10072110}.

\bibitem{Chen2011}
W.~C. Chen, J.~J. Mock, D.~R. Smith, D.~R. Akalin, and W.~J. Padilla,
  ``{Controlling Gigahertz and Terahertz Surface ElectromagneticWaves with
  Metamaterial Resonators},'' \emph{Physical Review X}, vol.~1, no.~2, p.
  021016, 12 2011, doi:
  \href{http://dx.doi.org/10.1103/PhysRevX.1.021016}{10.1103/PhysRevX.1.021016}.

\bibitem{Horestani2013a}
A.~K. Horestani, W.~Withayachumnankul, A.~Chahadih, A.~Ghaddar, M.~Zehar,
  D.~Abbott, C.~Fumeaux, and T.~Akalin, ``{Metamaterial-inspired bandpass
  filters for terahertz surface waves on goubau lines},'' \emph{IEEE
  Transactions on Terahertz Science and Technology}, vol.~3, no.~6, pp.
  851--858, 2013, doi:
  \href{http://dx.doi.org/10.1109/TTHZ.2013.2285556}{10.1109/TTHZ.2013.2285556}.

\bibitem{Young1966}
L.~Young, G.~L. Matthaei, and E.~M. Jones, ``{Microwave Band-Stop Filters With
  Narrow Stop Bands},'' \emph{IRE Transactions on Microwave Theory and
  Techniques}, vol. MTT-10, no.~6, pp. 416--427, 1962, doi:
  \href{http://dx.doi.org/10.1109/TMTT.1962.1125549}{10.1109/TMTT.1962.1125549}.

\bibitem{Marks1992}
R.~Marks and D.~Williams, ``{A general waveguide circuit theory},''
  \emph{Journal of Research of the National Institute of Standards and
  Technology}, vol.~97, no.~5, pp. 533--562, 1992, doi:
  \href{http://dx.doi.org/10.6028/jres.097.024}{10.6028/jres.097.024}.

\bibitem{Ribo2000}
M.~Rib{\'{o}} and L.~Pradell, ``{Circuit Model for a Coplanar-Slotline
  Cross},'' \emph{IEEE Microwave and Guided Wave Letters}, vol.~10, no.~12, pp.
  511--513, 2000, doi:
  \href{http://dx.doi.org/10.1109/75.895085}{10.1109/75.895085}.

\bibitem{Ponchak2018}
G.~E. Ponchak, ``{Coplanar Stripline Series-Stub with even Mode Suppression},''
  \emph{IEEE Microwave and Wireless Components Letters}, vol.~28, no.~11, pp.
  963--965, 2018, doi:
  \href{http://dx.doi.org/10.1109/LMWC.2018.2869283}{10.1109/LMWC.2018.2869283}.

\bibitem{Zelby1962}
L.~W. Zelby, ``{Propagation modes on a dielectric coated wire},'' \emph{Journal
  of the Franklin Institute}, vol. 274, no.~2, pp. 85--97, 8 1962, doi:
  \href{http://dx.doi.org/10.1016/0016-0032(62)90398-8}{10.1016/0016-0032(62)90398-8}.

\bibitem{Fikioris1979}
J.~G. Fikioris and J.~A. Roumeliotis, ``{Cutoff Wavenumbers of Goubau Lines},''
  \emph{IEEE Transactions on Microwave Theory and Techniques}, vol.~27, no.~6,
  pp. 570--573, 1979, doi:
  \href{http://dx.doi.org/10.1109/TMTT.1979.1129673}{10.1109/TMTT.1979.1129673}.

\bibitem{Dai2004}
J.~Dai, J.~Zhang, W.~Zhang, and D.~Grischkowsky, ``{Terahertz time-domain
  spectroscopy characterization of the far-infrared absorption and index of
  refraction of high-resistivity, float-zone silicon},'' \emph{Journal of the
  Optical Society of America B}, vol.~21, no.~7, p. 1379, 2004, doi:
  \href{http://dx.doi.org/10.1364/JOSAB.21.001379}{10.1364/JOSAB.21.001379}.

\bibitem{Rutledge1978}
D.~B. Rutledge, S.~E. Schwarz, and A.~T. Adams, ``{Infrared and submillimetre
  antennas},'' \emph{Infrared Physics}, vol.~18, no. 5-6, pp. 713--729, 1978,
  doi:
  \href{http://dx.doi.org/10.1016/0020-0891(78)90094-5}{10.1016/0020-0891(78)90094-5}.

\bibitem{Rutledge1983_thinCPW}
D.~B. Rutledge, D.~P. Neikirk, and D.~P. Kasilingam, ``{Integrated-Circuit
  Antennas},'' in \emph{Infrared and millimeter waves}.\hskip 1em plus 0.5em
  minus 0.4em\relax New York: Academic, 1983, ch.~1, pp. 59--63.

\bibitem{Grischkowsky1987}
D.~Grischkowsky, I.~N. Duling, J.~C. Chen, and C.~C. Chi, ``{Electromagnetic
  shock waves from transmission lines},'' \emph{Physical Review Letters},
  vol.~59, no.~15, pp. 1663--1666, 10 1987, doi:
  \href{http://dx.doi.org/10.1103/PhysRevLett.59.1663}{10.1103/PhysRevLett.59.1663}.

\bibitem{Hatkin1954}
L.~Hatkin, ``{Analysis of Propagating Modes in Dielectric Sheets},''
  \emph{Proceedings of the IRE}, vol.~42, no.~10, pp. 1565--1568, 1954, doi:
  \href{http://dx.doi.org/10.1109/JRPROC.1954.274764}{10.1109/JRPROC.1954.274764}.

\bibitem{Xu2006b}
Y.~Xu and R.~G. Bosisio, ``{A comprehensive study on the planar type of Goubau
  line for millimetre and submillimetre wave integrated circuits},'' \emph{IET
  Microwaves, Antennas and Propagation}, vol.~1, no.~3, pp. 681--687, 2007,
  doi:
  \href{http://dx.doi.org/10.1049/iet-map:20050308}{10.1049/iet-map:20050308}.

\bibitem{Harvey1960}
A.~F. Harvey, ``{Periodic and Guiding Structures at Microwave Frequencies},''
  \emph{IRE Transactions on Microwave Theory and Techniques}, vol.~8, no.~1,
  pp. 30--61, 1960, doi:
  \href{http://dx.doi.org/10.1109/TMTT.1960.1124658}{10.1109/TMTT.1960.1124658}.

\bibitem{Schelkunoff1944}
S.~A. Schelkunoff, ``{Impedance concept in wave guides},'' \emph{Quarterly of
  Applied Mathematics}, vol.~II, pp. 1--15, 1944, doi:
  \href{http://dx.doi.org/10.1090/qam/11833}{10.1090/qam/11833}.

\bibitem{Cavallo2017}
D.~Cavallo, W.~H. Syed, and A.~Neto, ``{Equivalent Transmission Line Models for
  the Analysis of Edge Effects in Finite Connected and Tightly Coupled
  Arrays},'' \emph{IEEE Transactions on Antennas and Propagation}, vol.~65,
  no.~4, pp. 1788--1796, 2017, doi:
  \href{http://dx.doi.org/10.1109/TAP.2017.2670616}{10.1109/TAP.2017.2670616}.

\bibitem{Cabello-Sanchez2018}
J.~Cabello-S{\'{a}}nchez, H.~Rodilla, V.~Drakinskiy, and J.~Stake, ``{Multiline
  TRL Calibration Standards for S-parameter Measurement of Planar Goubau Lines
  from 0.75 THz to 1.1 THz},'' in \emph{2018 IEEE/MTT-S International Microwave
  Symposium - IMS}, Philadelphia, PA, USA, 2018, pp. 879--882, doi:
  \href{http://dx.doi.org/10.1109/MWSYM.2018.8439138}{10.1109/MWSYM.2018.8439138}.

\bibitem{Cabello-Sanchez2021}
J.~Cabello-Sanchez, V.~Drakinskiy, J.~Stake, and H.~Rodilla, ``{On-Chip
  Characterization of High-Loss Liquids between 750 and 1100 GHz},'' \emph{IEEE
  Transactions on Terahertz Science and Technology}, vol.~11, no.~1, pp.
  113--116, 2021, doi:
  \href{http://dx.doi.org/10.1109/TTHZ.2020.3029503}{10.1109/TTHZ.2020.3029503}.

\bibitem{Bauwens2014}
M.~F. Bauwens, N.~Alijabbari, A.~W. Lichtenberger, N.~S. Barker, and R.~M.
  Weikle, ``{A 1.1 THz micromachined on-wafer probe},'' in \emph{2014 IEEE
  MTT-S International Microwave Symposium (IMS2014)}, Tampa, FL, 2014, pp.
  1--4, doi:
  \href{http://dx.doi.org/10.1109/MWSYM.2014.6848607}{10.1109/MWSYM.2014.6848607}.

\bibitem{Fuse2011}
N.~Fuse, T.~Takahashi, Y.~Ohki, R.~Sato, M.~Mizuno, and K.~Fukunaga,
  ``{Terahertz Spectroscopy as a New Tool for Insulating Material Analysis and
  Condition Monitoring},'' \emph{IEEE Insulating Magazine}, vol.~27, no.~3, pp.
  26--35, 2011, doi:
  \href{http://dx.doi.org/10.1109/MEI.2011.5871366}{10.1109/MEI.2011.5871366}.

\bibitem{Holloway2020}
J.~W. Holloway, G.~C. Dogiamis, S.~Shin, and R.~Han, ``{220-to-330-GHz manifold
  triplexer with wide stopband utilizing ridged substrate integrated
  waveguides},'' \emph{IEEE Transactions on Microwave Theory and Techniques},
  vol.~68, no.~8, pp. 3428--3438, 2020, doi:
  \href{http://dx.doi.org/10.1109/TMTT.2020.2997367}{10.1109/TMTT.2020.2997367}.

\bibitem{Wang2020}
F.~Wang, V.~F. Pavlidis, and N.~Yu, ``{Miniaturized SIW Bandpass Filter Based
  on TSV Technology for THz Applications},'' \emph{IEEE Transactions on
  Terahertz Science and Technology}, vol.~10, no.~4, pp. 423--426, 2020, doi:
  \href{http://dx.doi.org/10.1109/TTHZ.2020.2974091}{10.1109/TTHZ.2020.2974091}.

\bibitem{Ding2021}
J.~Q. Ding, J.~Hu, and S.~C. Shi, ``{350-GHz Bandpass Filters Using
  Superconducting Coplanar Waveguide},'' \emph{IEEE Transactions on Terahertz
  Science and Technology}, vol.~11, no.~5, pp. 548--556, 2021, doi:
  \href{http://dx.doi.org/10.1109/TTHZ.2021.3071019}{10.1109/TTHZ.2021.3071019}.

\bibitem{PascualLaguna2021}
A.~Pascual~Laguna, K.~Karatsu, D.~Thoen, V.~Murugesan, B.~Buijtendorp, A.~Endo,
  and J.~Baselmans, ``{Terahertz Band-Pass Filters for Wideband Superconducting
  On-Chip Filter-Bank Spectrometers},'' \emph{IEEE Transactions on Terahertz
  Science and Technology}, vol.~11, no.~6, pp. 635--646, 2021, doi:
  \href{http://dx.doi.org/10.1109/TTHZ.2021.3095429}{10.1109/TTHZ.2021.3095429}.

\bibitem{Cunningham2005}
J.~Cunningham, C.~Wood, A.~G. Davies, I.~Hunter, E.~H. Linfield, and H.~E.
  Beere, ``{Terahertz frequency range band-stop filters},'' \emph{Applied
  Physics Letters}, vol.~86, no.~21, pp. 1--3, 2005, doi:
  \href{http://dx.doi.org/10.1063/1.1938255}{10.1063/1.1938255}.

\end{thebibliography}


\begin{thebibliography}{1}

\bibitem{IEEEhowto:kopka}
H.~Kopka and P.~W. Daly, \emph{A Guide to \LaTeX}, 3rd~ed.\hskip 1em plus
  0.5em minus 0.4em\relax Harlow, England: Addison-Wesley, 1999.

\end{thebibliography}

\begin{IEEEbiography}[{\includegraphics[width=1in,height=1.25in,clip,keepaspectratio]{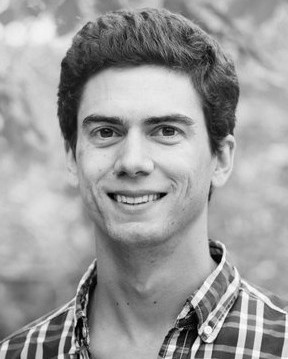}}]{Juan Cabello-S\'{a}nchez}
(S' 17) was born in Madrid, Spain in 1992. He received the Bachelor diploma in Electrical Engineering and Computer Science from the Technical University of Madrid, Spain, in 2015, and MSc in Wireless, Photonics and Space Engineering from Chalmers University of Technology, in Gothenburg, Sweden, in 2017.

From July 2017 he is pursuing his Ph.D. in the Terahertz and Millimetre-wave Laboratory at Chalmers University of Technology.
\end{IEEEbiography}

\begin{IEEEbiography}[{\includegraphics[width=1in,height=1.25in,clip,keepaspectratio]{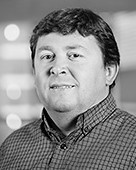}}]{Vladimir Drakinskiy}
Vladimir Drakinskiy was born in Kurganinsk, Russia, in 1977. He received the Diploma degree in physics and informatics (with honors) from the Armavir State Pedagogical Institute, Armavir, Russia, in 2000, and the Postgraduate degree from Moscow State Pedagogical University, Moscow, Russia, in 2003.

From 2000 to 2003, he was a Junior Research Assistant in the Physics Department, Moscow State Pedagogical University. Since 2003, he has been in the Department of Microtechnology and Nanoscience, Chalmers University of Technology, Gothenburg, Sweden. During 2003–2005, he was responsible for mixer chips fabrication for the Herschel Space Observatory. Since 2008, he has been a Research Engineer with the Department of Microtechnology and Nanoscience, Chalmers University of Technology. He is currently responsible for terahertz Schottky diodes process line at MC2, Chalmers University of Technology. His research interests include microfabrication and nanofabrication techniques, detectors for submillimeter and terahertz ranges, and superconducting thin films.
\end{IEEEbiography}

\begin{IEEEbiography}[{\includegraphics[width=1in,height=1.25in,clip,keepaspectratio]{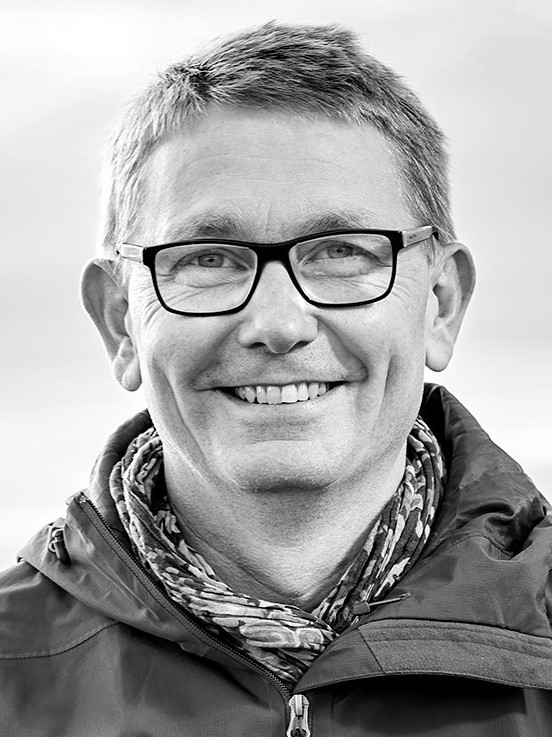}}]{Jan Stake}
(S' 95--M' 00--SM' 06) was born in Uddevalla, Sweden, in 1971. He received the M.Sc. degree in electrical engineering and the Ph.D. degree in microwave electronics from the Chalmers University of Technology, Goteborg, Sweden, in 1994 and 1999, respectively.

In 1997 he was a Research Assistant with the University of Virginia, Charlottesville, USA. From 1999 to 2001, he was a Research Fellow with the Millimetre Wave Group at the Rutherford Appleton Laboratory, Didcot, UK. He then joined Saab Combitech Systems AB, as a Senior RF/microwave Engineer, until 2003. From 2000 to 2006, he held different academic positions with Chalmers University of Technology and, from 2003-2006, was also Head of the Nanofabrication Laboratory, Department of Microtechnology and Nanoscience (MC2). During 2007, he was a Visiting Professor with the Submillimeter Wave Advanced Technology (SWAT) Group at Caltech/JPL, Pasadena, USA. In 2020, he was a Visiting Professor at TU Delft. He is currently Professor and Head of the Terahertz and Millimetre Wave Laboratory, Chalmers University of Technology, Sweden. He is also cofounder of Wasa Millimeter Wave AB, Göteborg, Sweden. His research involves graphene electronics, high-frequency semiconductor devices, THz electronics, submillimeter wave measurement techniques, and terahertz systems.

Prof. Stake served as Editor-in-Chief for the IEEE TRANSACTIONS ON TERAHERTZ SCIENCE AND TECHNOLOGY between 2016-2018 and topical editor between 2012-2015.
\end{IEEEbiography}

\begin{IEEEbiography}[{\includegraphics[width=1in,height=1.25in,clip,keepaspectratio]{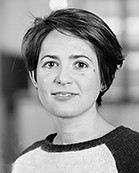}}]{Helena Rodilla}
(M' 16--SM' 20) was born in Salamanca, Spain, in 1982. She received the B.S. and Ph.D. degrees in Physics from the University of Salamanca, Salamanca, Spain, in 2006 and 2010, respectively.

From 2006 to 2010, she was with the Electronics Group, Department of Applied Physics, University of Salamanca, Spain, where her research interest was semiconductor physics. From 2011 to 2013 she was Postdoctoral Researcher with the Microwave Electronics Laboratory, Department of Microtechnology and Nanoscience (MC2), Chalmers University of Technology, Gothenburg, Sweden, where she worked on very low-noise InP HEMTs for cryogenic low noise amplifiers. Since 2013 she has been with the Terahertz and Millimetre Wave Laboratory, MC2, Chalmers University of Technology, Gothenburg, Sweden, where she became Associate Professor in 2020. Her current research interests include the use of millimeter wave and terahertz technology in life science applications, sensing and on-wafer terahertz measurements.

\end{IEEEbiography}






\end{document}